\documentclass{jfm}
\usepackage{bm}
\usepackage{natbib}
\usepackage{epsfig}
\usepackage{amssymb}
\usepackage[dvips]{color}
\usepackage{lscape}
\usepackage{longtable}
\usepackage{rotating}
\usepackage{amsmath}
\usepackage{graphicx}

\title[On the lateral migration of a slightly deformed bubble]
{On the lateral migration of a slightly deformed bubble rising near a vertical plane wall}

\author[K. Sugiyama and F. Takemura]
{KAZUYASU SUGIYAMA$^1$ and FUMIO TAKEMURA$^{2, 1}$
}

\affiliation{
$^1$Department of Mechanical Engineering, 
School of Engineering, The University of Tokyo,
7-3-1 Hongo, Bunkyo-ku, Tokyo 113-8656, Japan\\
$^2$National Institute of Advanced Industrial Science and Technology, 
1-2-1 Namiki, Tsukuba, Ibaraki 305-8564, Japan
}

\date{\today}
\begin{document}

\maketitle

\begin{abstract}
Deformation-induced lateral migration of a bubble slowly rising near a vertical
plane wall in a stagnant liquid is numerically and theoretically investigated.
In particular, our focus is set on a situation with a short clearance $c$
between the bubble interface and the wall.
Motivated by the fact that numerically 
and experimentally 
measured migration velocities 
are considerably higher than the velocity estimated by the available analytical solution
using the Fax\'{e}n mirror image technique 
for $a/(a+c)\ll 1$ (here $a$ is the bubble radius), 
when the clearance parameter $\varepsilon(= c/a)$ 
is comparable to or smaller than unity, 
the numerical analysis based on 
the boundary-fitted finite-difference approach solving the Stokes equation
is performed to complement the experiment.
The migration velocity is found to be more affected by the high-order deformation modes
 with decreasing $\varepsilon$.
The numerical simulations are compared with 
 a theoretical migration velocity obtained from 
 a lubrication study of a nearly spherical drop,
 which describes the role of the squeezing flow within the bubble-wall gap.
The numerical and lubrication analyses consistently demonstrate
that when $\varepsilon\leq 1$, the lubrication effect
 makes the migration velocity asymptotically
$\mu V_{B1}^2/(25\varepsilon \gamma)$
(here, $V_{B1}$, $\mu$, and $\gamma$ denote the rising velocity, 
 the dynamic viscosity of liquid, 
 and the surface tension, respectively).

\end{abstract}

\section{Introduction}

Recent technical progress in generating microbubbles 
(e.g. \cite{gar2006, mak2006}),
including potentials as actuator and sensor, 
has enhanced the range of applications, 
e.g. additives to reduce a turbulence friction (\cite{ser2005}), 
drug delivery capsules (\cite{sho2004}), and
contrast agents (\cite{cor2001}).
In many situations, a bubble encounters a boundary wall 
during its transport process, 
and a hydrodynamic interaction occurs as characterized by the
inter-scale between the bubble and the wall.
In practice, it is of primary importance that 
the bubble undergoes a repulsive or attractive force in the wall-normal direction, 
which causes a lateral migration (\cite{lea1980, mag2003, hib2007})
and determines the bubble distribution, 
when translating parallel to the wall.
As the simplest model system, 
one may raise a phenomenon of a spherical bubble rising near a vertical
infinite plane wall in a creeping (Stokes) flow.
However, there is no mechanism to generate the lateral migration force, 
 as kinematic reversibility is ensured by
 symmetry of the boundary and 
by linearity in the Stokes equation (\cite{lea1992}, chapter 4). 
In fact, the migration force stems from nonlinearities in 
the advective momentum transport 
(\cite{cox1968, ho1974, vas1976, vas1977, cox1977, mcl1993, che1994, bec1996, mag2003})
and/or the interfacial deformability 
(\cite{cha1965, cha1979, sha1988, uij1993, uij1995, mag2003, wan2006}),
to break the symmetry.

For a tank-treading vesicle translating parallel to the wall, 
 the migration force was theoretically obtained by \cite{oll1997}, 
 in which the vesicle shape is prescribed as a strongly non-spherical ellipsoid, 
 and the theory was experimentally validated by \cite{cal2008}.
For a bubble or drop, the shape cannot be prescribed
 since it obeys the Laplace law and depends on the surrounding fluid flow.
The theoretical success in solving the nontrivial problem of the deformation-induced
 migration of the bubble or drop was made by \cite{mag2003} 
 using the Fax\'{e}n mirror image technique and the Lorentz reciprocal theorem.
However, \cite{wan2006} performed a numerical study 
 on the motion of a drop with the same viscosity as the surrounding 
 fluid in a linear shear flow by means of a boundary element
 method, and pointed out that the theory considerably underestimates the
 migration velocity or erroneously predicts the lateral motion, 
 despite consistent predictions of the rising velocity and the interfacial deformation. 
Recently, \cite{tak2009} experimentally measured 
 the lateral migration velocity of slightly deformed bubbles 
 in a wall-bounded shear flow, 
 and found a clear discrepancy between 
 the experimental and theoretical values 
 of the deformation-induced transverse force. 
Then, they computed the quasi-steady evolution of deformable bubbles moving
 in a wall-bounded linear shear flow at zero Reynolds number using
 a spectral boundary element method developed by \cite{dim2007},
 and found that the measured deformation-induced lift force agrees
 quantitatively well with the computational prediction. 
Motivated by their conclusions, we revisited experimental data of the bubble migration 
 in a quiescent liquid obtained by \cite{tak2002}, and analysed the 
 data on the conditions
 that the clearance between the bubble interface and the wall
 is comparable to or shorter than the bubble radius, 
 which was not considered there. 
The results revealed that the discrepancy between the migration 
 velocities of the experiment and the theory
 increased as the bubble was closed to the wall, as detailed below.

\begin{figure}
\begin{center}

\vspace{1em}

\epsfig{file=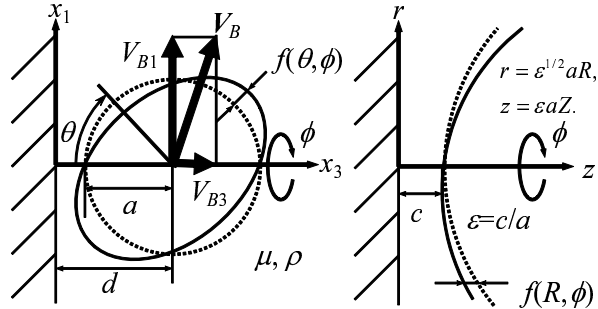,width=8cm,angle=0}
\end{center}
\caption{
Schematic of a buoyant bubble moving near a vertical plane wall in a quiescent liquid. 
The solid outline shows a deformed interface with 
a deflection $f$ from a spherical interface indicated by the dotted outline. 
(a) left panel: coordinates system around the bubble. 
In the analysis of \cite{mag2003},
 $a/d$ is assumed to be sufficiently smaller than unity
 (here, $a$ is the bubble radius and $d$ is the distance between the bubble centroid and the wall).
(b) right panel: inter-scale coordinates between the bubble surface and the wall 
and scaling relations suitable to a lubrication theory. 
A clearance parameter is defined as $\varepsilon=c/a$.
}
\label{fig:schem}
\end{figure}

\begin{figure}
\begin{center}

\epsfig{file=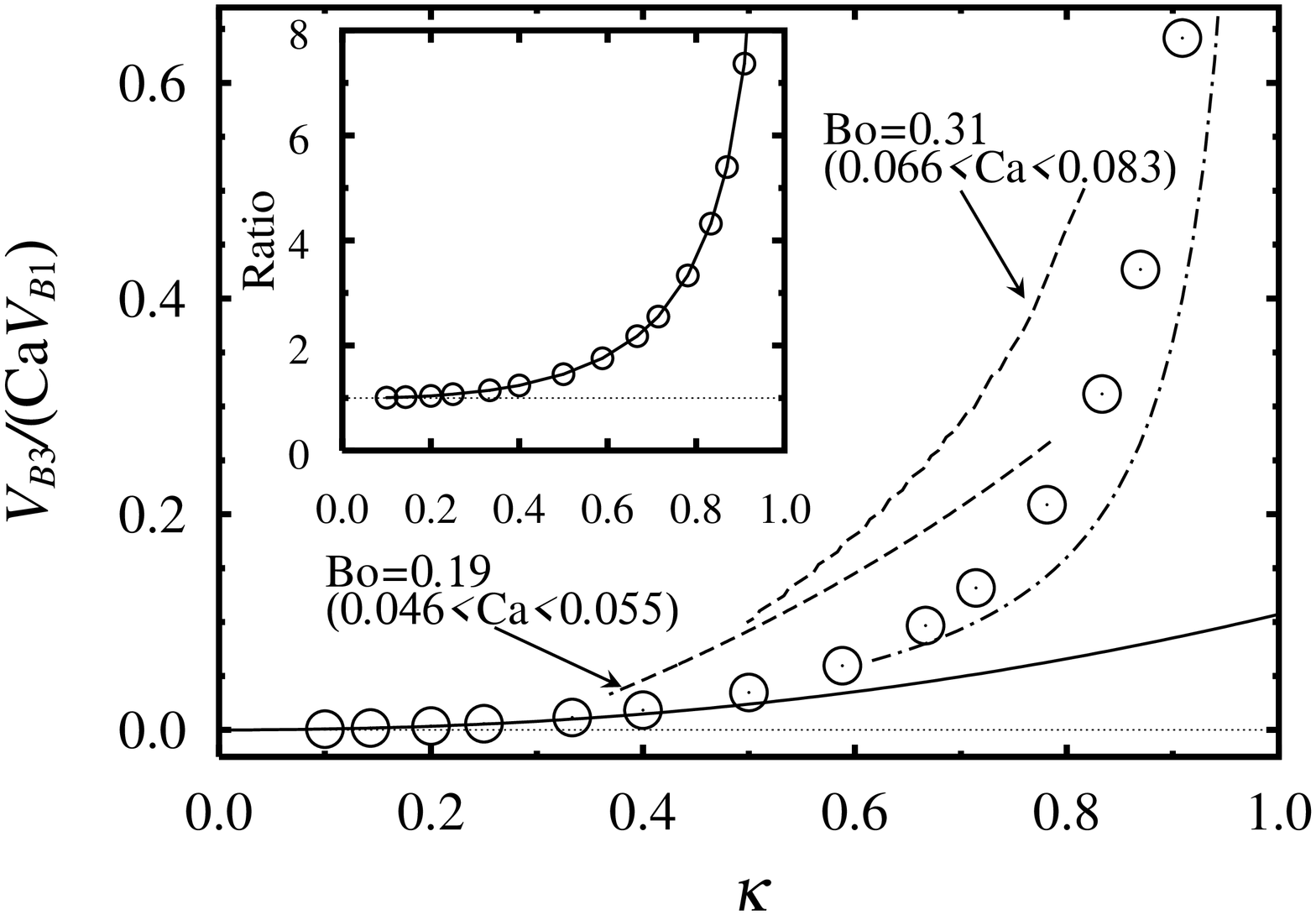,width=7cm,angle=270}

\end{center}
\caption{\small
Migration velocity $V_{B3}$ versus $\kappa (=(1+\varepsilon)^{-1})$. 
The solid curve indicates the analytical solution (\cite{mag2003})
$V_{B3}/({\rm Ca}V_{B1})$
$=3\kappa^2$
$(1+3\kappa/2)$
$/\{40(1+3\kappa/4)\}$
with the assumption of 
the sufficiently long distance between 
the bubble centroid and the wall, i.e. $\varepsilon\gg 1$.
The dashed curves indicate the experimental results (\cite{tak2002}), 
 the circles the results obtained by the numerical simulation,
 and the dashed-dotted curve the prediction (\ref{eq:vb3_lub}) 
 by means of the lubrication approach (\cite{hod2004}). 
The inset shows the ratio of the simulation result to the analytical solution.
}
\label{fig:migvel_comp_exp}
\end{figure}

In this paper, we focus on the bubble motion in a quiescent liquid to simplify the subject. 
Let us consider the migration velocity $V_{B3}(={\bm V}_{B}\cdot {\bm e}_{3})$
 of a bubble rising near a vertical plane wall at a distance $d$ between the bubble centroid and the wall
 as schematically illustrated in figure \ref{fig:schem}(a).
The bubble has an equivalent radius $a$ to that of a sphere with the same volume. 
Introducing an interfacial deflection $f(\theta,\phi)$ from a sphere,
 we write the distance from the bubble centroid to the interface as $a+f$.
The experimental results used here were measured under the condition that
 the Reynolds number ${\rm Re}=2\rho aV_{B1}/\mu$ 
 (here $V_{B1}(={\bm V}_B\cdot{\bm e}_{1}$), 
 $\rho$, and $\mu$ respectively denote 
 the rising velocity, the density, and the dynamic viscosity of liquid) 
 is unity or less (\cite{tak2002}). 
The pure lateral migration velocities induced by the deformation $V_{B3}$
 were calculated from the measured values substituting the velocities induced 
 by the inertia effects. 
 Following a Stokes flow theory for the deformation-induced migration (\cite{mag2003}), 
 we can characterize the system using two parameters, 
 i.e., a clearance parameter $\varepsilon(= c/a)$ 
 and a capillary number ${\rm Ca}=\mu V_{B1}/\gamma$ 
(or a Bond number ${\rm Bo}=\rho a^2 g/\gamma$ as used in \cite{mag2003}).
Here $\gamma$ and $g$ respectively denote
 the surface tension and the acceleration of gravity. 
Further, as long as ${\rm Ca}\ll 1$, 
 we may use ${\rm Ca}$ as a perturbation parameter, 
 and reduce the ${\rm Ca}$-dependent system to another, 
 in which $V_{B1}$, $V_{B3}/{\rm Ca}$ and $f/{\rm Ca}$
 are dependent only upon $\varepsilon$, under the infinitesimal deformation assumption.
Figure \ref{fig:migvel_comp_exp} shows the migration velocity $V_{B3}$ 
 away from the wall 
 normalized by ${\rm Ca}V_{B1}$ 
 as a function of $\kappa (\equiv (1+\varepsilon)^{-1})$.
(It should be noticed that
 although $\kappa$ as well as $\varepsilon$
 are measures of the distance between the wall and the bubble,
 hereafter $\kappa$ is also used to make some equations for the wide gap case ($\kappa\ll 1$) simple.)
The measured velocity is 
 found to be much higher than the analytical
 solution especially for the large $\kappa$.
A possible inference drawn from this result is that 
 there exists an additional ingredient
 to generate repulsive force for narrow bubble-wall gap, 
 which is not covered by the theory of \cite{mag2003}.

As the most crucial restriction involved in the mirror image technique, 
we can raise an assumption that the bubble-wall distance is much longer than the bubble radius. 
However, with regard to an inertia effect on the lateral velocity of a rigid sphere,
\cite{tak2004} experimentally demonstrated that the mirror image approach
has robust applicability in prediction beyond the wide-gap precondition. 
At this moment, we cannot conclude whether the discrepancy in 
the deformation-induced migration velocity comes from 
the erroneous prediction due to the miscalculation
or the contradictory conditions in the boundary element computation and the experiment with the theoretical assumptions.
To complement the experiment and to gain further insight into the deformation-induced migration,
we investigate the migration behavior with attention to shortness of a bubble-wall clearance $c(=d-a)$.
As in \cite{mag2003}, 
using two Stokes flow solutions for a spherical bubble translating parallel
and perpendicular to the wall,
we apply the Lorentz reciprocal theorem to evaluating the migration velocity.
We carry out numerical simulations using a boundary-fitted grid,
which can accurately implement the boundary conditions
and release the constraint of the sufficiently wide bubble-wall gap in the mirror image technique.
In addition, 
 comparisons with a theoretical migration velocity for $\varepsilon\ll 1$ obtained from 
 a lubrication study of a nearly spherical drop moving near a tilted plane (\cite{hod2004}),
 in which the secondary flow due to the change in the boundary geometry caused by the bubble deformation
 is responsible for the wall-normal force,
 are made to shed more light on the short clearance effect.

\section{Numerical simulation}

\subsection{General formulation}
\label{sec:gf}

To clarify the physical mechanism of the repulsive force, 
we numerically address the bubble migration. 
In a similar manner to \cite{mag2003}, 
instead of directly solving the flow field with the deformed bubble, 
we employ the Lorentz reciprocal theorem 
to determine the lateral migration force and velocity
through coupling two flow fields around a spherical bubble
translating parallel and perpendicular to the wall. 
In the subsequent developments,
 the basic equations and the involved variables are nondimensionalized 
 using $a$, $V_{B1}$ and $\mu$.
We suppose that the bubble quasi-steadily rises near an infinite flat plate
in a stagnant incompressible liquid, 
and both the Reynolds and capillary numbers
are sufficiently smaller than unity. 
Hence, the system is described by the steady Stokes equation 
for solenoidal velocity vectors, i.e.
\begin{equation}
\nabla\cdot{\bm U}=\nabla\cdot{\bm u}=0,
\ \ \ -\infty<x_1<\infty,\ -\infty<x_2<\infty,\ -1-\varepsilon\le x_3<\infty,
\label{eq:cont_st}
\end{equation}
\begin{equation}
\nabla\cdot{\bm \Sigma}=\nabla\cdot{\bm \sigma}=0,
\ \ \ -\infty<x_1<\infty,\ -\infty<x_2<\infty,\ -1-\varepsilon\le x_3<\infty,
\label{eq:mom_st}
\end{equation}
where $({\bm U},{\bm \Sigma})$ and $({\bm u},{\bm \sigma})$ 
are the velocity and stress fields for the bubble translating
respectively parallel and perpendicular at a speed of unity to the wall.
The ${\rm Ca}$ dependence of the interfacial deflection is given by
$f(\theta,\phi;{\rm Ca})\ ={\rm Ca}f^{({\rm Ca})}(\theta,\phi)$.
The bubble deformation obeys the Laplace law
for the infinitesimal deflection $|f|\ll 1$ with ${\rm Ca}\ll 1$, 
\begin{equation}
\left(\nabla_s^2+2\right)f^{({\rm Ca})}=
-{\bm n}\cdot{\bm\Sigma}\cdot{\bm n}
+3x_1 \langle x_1{\bm n}\cdot{\bm \Sigma}\cdot{\bm n}\rangle_{S_B}
\ \ {\rm at}\ \ \sqrt{x_1^2+x_2^2+x_3^2}=1,
\label{eq:laplace_law}
\end{equation}
where 
${\bm n}$ represents the normal unit vector pointing outwards the liquid,
$\nabla_s(=\nabla-{\bm n}({\bm n}\cdot\nabla))$ is the nabla operator along the tangential directions
on the bubble surface, 
$\langle ...\rangle_{S_B}$ is the area average taken over the bubble surface, 
and $x_1$ is the coordinates in the upward direction from the origin at the bubble centroid.
Kinematic and free-slip conditions are imposed on the bubble surface, i.e.
\begin{equation}
\left.
\begin{array}{rl}
{\bm n}\cdot{\bm U}&=0,\\
({\bm n}\cdot{\bm\Sigma})\times{\bm n}&=0,\\
{\bm n}\cdot{\bm u}&=0,\\
({\bm n}\cdot{\bm\sigma})\times{\bm n}&=0,\\
\end{array}
\right\}
\ \ {\rm at}\ \ \sqrt{x_1^2+x_2^2+x_3^2}=1,
\label{eq:bc_bubsurf_st}
\end{equation}
where we take the reference frames viewed from the bubble. 
On the plane wall, we impose the no-slip condition 
\begin{equation}
\left.
\begin{array}{rl}
{\bm U}&=-{\bm e}_1,\\
{\bm u}&={\bm e}_3,
\end{array}
\right\}
\ \ {\rm at}\ \ x_3=-1-\varepsilon.
\label{eq:wall_st}
\end{equation}
Sufficiently far from the bubble, the velocity vectors
approach the uniform velocities
\begin{equation}
\left.
\begin{array}{rl}
{\bm U}&\rightarrow -{\bm e}_1,\\
{\bm u}&\rightarrow {\bm e}_3,\\
\end{array}
\right\}
\ \ {\rm as}\ \ \sqrt{x_1^2+x_2^2+x_3^2}\rightarrow \infty.
\label{eq:far_st}
\end{equation}
Thanks to the reciprocal theorem (\cite{lea1980}; (35) of \cite{mag2003}), 
the deformation-induced lateral force $F_M={\rm Ca}F_M^{({\rm Ca})}$ 
to cancel the migration velocity and to maintain the wall-parallel motion is expressed as
\begin{equation}
F_{M}^{({\rm Ca})}=
\oint_{S_B}\!\!\!\!\!{\rm d}^2{\bm x}\ 
{\cal L}(f^{({\rm Ca})}),
\label{eq:fmca_st}
\end{equation}
where $S_B$ denotes the bubble surface,
and the operator ${\cal L}$ is given by 
\begin{equation}
\begin{split}
&{\cal L}=
{\bm n}\!\cdot{\bm \sigma}\!\cdot{\bm n}
\left(
\frac{\partial{\bm U}}{\partial n}\!\cdot\!{\bm n}-{\bm U}\!\cdot\!\nabla_s
\right)
-{\bm u}\!\cdot\!\biggl(
\frac{\partial {\bm \Sigma}}{\partial n}\!\cdot\!{\bm n}
-{\bm \Sigma}\!\cdot\!\nabla_s
\biggr)
\\&
-\left\{
{\bm n}\cdot{\bm\Sigma}\cdot{\bm n}
-3x_1\langle x_1 {\bm n}\cdot{\bm\Sigma}\cdot{\bm n}\rangle_{S_B}
\right\}{\bm u}\cdot\nabla_s.
\end{split}\end{equation}
The migration velocity $V_{B3}={\rm Ca}V_{B3}^{({\rm Ca})}$ is expressed as
\begin{equation}
\frac{V_{B3}^{({\rm Ca})}}{V_{B1}}=\frac{F_{M}^{({\rm Ca})}}{F_{DC}},
\label{eq:vb3ca}
\end{equation}
where \begin{equation}
F_{DC}=\oint_{S_B}\!\!\!{\rm d}^2{\bm x}\ {\bm e}_3\cdot{\bm\sigma}\cdot{\bm n}
\label{eq:fdc}
\end{equation}
denotes the drag force acting on the bubble translating perpendicular to
the plane wall.

\subsection{Simulation method}

The basic equations are numerically solved 
by the second-order finite-difference method discretized on the bipolar
coordinates $(\xi,\eta)$ grid,
which is boundary-fitted on both the bubble surface and the plane wall
(see figure \ref{fig:schem_bipolar}(a) in Appendix \ref{appendix_a}).
We take care of the mass and momentum conservations in a discretized form. 
For technical detail on the discretization, see Appendix \ref{appendix_a}.
The number of grid points is $N_\xi\times N_\eta = 200 \times 200$, 
and the grid is non-uniform and refined near the wall and the bubble surface.
The computational procedure is based on 
a Simplified-Marker-And-Cell method (\cite{ams1970})
with a first-order Eulerian implicit time marching scheme. 
Such an unsteady scheme enables us to 
 check whether the computation converges to 
 the fully developed state through temporal changes in
 the budgets of the momentum and kinetic-energy transports.
To avoid a problem associated with singularities
 in the discretization near the axis, 
 we follow a method proposed by \cite{fuk2002}.

For each run, we confirm that drag forces, numerically evaluated on the bubble surface, 
 for both the perpendicular and parallel motions 
 are in good agreement with the respective kinetic-energy dissipation rates, 
 numerically integrated over the entire computational domain,
 normalized by the translational velocities 
 with an error of less than $0.040\%$. 
Such an agreement between the surface and bulk quantities
 indicates that the computation 
 is well converged and reaches to the steady state 
 in view of the momentum and kinetic-energy budgets.
Further, the drag force for the perpendicular motion 
shows quantitative agreement with the infinite series solution 
of \cite{bar1968} with an error of less than $0.043\%$.
The drag force for the parallel motion approaches 
the wide-gap solution of \cite{mag2003} with increasing $\varepsilon$.
To make sure of numerical stability and accuracy, 
we set the clearance parameter in a range of $10^{-3}\leq \varepsilon\leq 9$.
We performed the convergence tests
 with varying the size $R_{\rm max}$ of the computational domain, 
 the number $N_\xi$ of nodes in the gap between the bubble and the wall,
 and the number $N_\eta$ of nodes describing the bubble surface. 
We confirmed that the relative errors in the migration velocity $V_{B3}$
 and the drag forces for the perpendicular and parallel bubble motions
 to those obtained on the base meshes $N_\xi\times N_\eta = 200 \times 200$ for various $\varepsilon$
 decrease with increasing the size $R_{\rm max}$ and 
 the resolutions $N_\xi$ and $N_\eta$. 
From the convergence behaviors, we deduce to obtain the migration velocities $V_{B3}$ 
 with errors of much less than $1\%$ using the present base meshes, 
 which are accurate enough for the subsequent discussion.

\subsection{Migration velocity}

As shown in the inset of figure \ref{fig:migvel_comp_exp}, 
 the ratio of the simulation to the analytical migration velocity
 becomes close to unity as $\kappa (=(1+\varepsilon)^{-1})$ approaches zero, 
 indicating the simulation result is consistent with
 the analytical solution (\cite{mag2003}) for small $\kappa (< \sim 0.3)$.
Such a consistency between the different approaches 
 may deny the erroneous prediction of
 the migration velocity in \cite{mag2003}, of which the possibility
 was pointed out by \cite{wan2006}.
By contrast, the simulated migration velocity for the bubble closer to the wall 
 with the clearance shorter than the bubble radius
(i.e., for $\kappa > \sim 0.5$ presumably beyond the theoretical precondition $\kappa\ll 1$) 
 is considerably higher than the analytical solution.
Although the simulation result reveals lower velocity than the experimental one,
 the tendency of the higher velocity than the theoretical one for the narrow gap 
 is qualitatively similar to that in the experiment.
Thus, the present simulation also indicates 
 the presence of the additional narrow-gap repulsive force.

\begin{figure}
\begin{center}

\epsfig{file=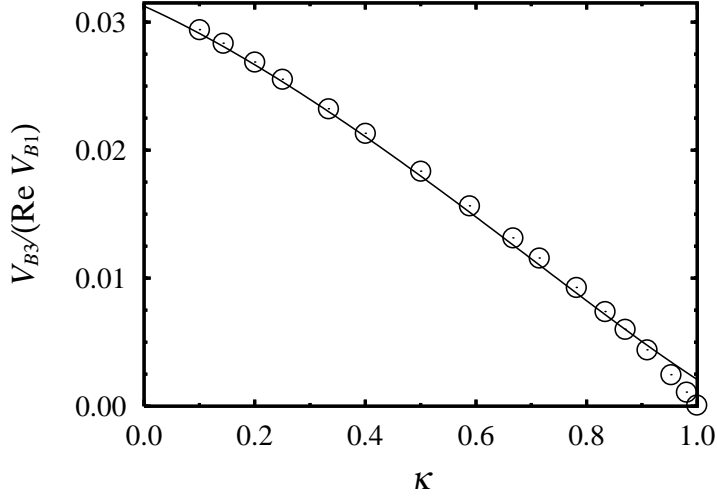,height=10cm,angle=270}

\end{center}
\caption{
Inertia-driven migration velocity $V_{B3}$ versus $\kappa (=(1+\varepsilon)^{-1})$ for spherical bubble at $0<{\rm Re}\ll 1$.
The solid curve indicates the analytical solution (\cite{mag2003})
$V_{B3}/({\rm Re}V_{B1})
=\frac{1}{32}(1+\frac{1}{8}\kappa
-0.516\kappa^2)(1-\frac{3}{4}\kappa-\frac{9}{64}\kappa^4)$
under the assumption of ($O({\rm Re})< \kappa \ll 1$) as in figure \ref{fig:migvel_comp_exp}.
The circles indicate the results obtained by the numerical simulation.
}
\label{fig:migvel_inert}
\end{figure}

For an undeformed spherical bubble at small but non-zero Reynolds
numbers $0<{\rm Re}\ll 1$,
using the solutions to the equation set (\ref{eq:cont_st})--(\ref{eq:far_st}),
we can also evaluate the inertia effect on the migration velocity 
$V_{B3}={\rm Re}\ V_{B3}^{({\rm Re})}$ from 
\begin{equation}
F_M^{({\rm Re})}=
\frac{1}{2}
\int_{{\cal V}}\!\!\!{\rm d}^3{\bm x}\ 
({\bm e}_3-{\bm u})\cdot\{({\bm U}\cdot\nabla){\bm U}\},
\label{eq:fmre_st}
\end{equation}
\begin{equation}
\frac{V_{B3}^{({\rm Re})}}{V_{B1}}=\frac{F_M^{({\rm Re})}}{F_{DC}},
\label{eq:vb3re_st}
\end{equation}
where ${\cal V}$ stands for the entire volume of liquid around the bubble.
The force expression (\ref{eq:fmre_st}) is theoretically justified for
the case that the wall is placed within a Stokes expansion region,
i.e., $O({\rm Re})< \kappa$ (\cite{vas1976}).
Figure \ref{fig:migvel_inert} shows the inertia-driven migration
velocity as a function of $\kappa$. 
The simulated profile is globally consistent with the analytical solution (\cite{mag2003})
even for the narrow gap $\kappa\sim 1$,
as opposed to the profile of the deformation-induced migration velocity in
figure \ref{fig:migvel_comp_exp}.
It should be noticed that 
the expression (\ref{eq:fmca_st}) of the deformation-induced lateral force 
is written in a surface integral form, 
while the inertia-driven force (\ref{eq:fmre_st}) in a volume integral form. 
The overall agreement in the inertia-driven migration velocity indicates that
capturing the bulk velocity distributions is important for predicting the migration velocity
and can be robustly attained by the mirror image technique over the wide range of $\kappa$. 
The wide range agreement with the theories (\cite{vas1976, mag2003})
was also experimentally demonstrated for the sedimenting rigid particle in a
range of $0.1<{\rm Re}< 1$ by \cite{tak2004}
as long as the wall is placed in the Stokes expansion region.
Contrastingly, as shown in figure \ref{fig:migvel_comp_exp}, 
the larger discrepancy between the deformation-induced migration velocities
of the simulation and the theory with increasing $\kappa$ 
indicates that the migration velocity is sensitive to 
the local effect leading to the additional narrow-gap repulsive force, 
which may not be covered by the mirror image technique.

\begin{figure}
\begin{center}

\epsfig{file=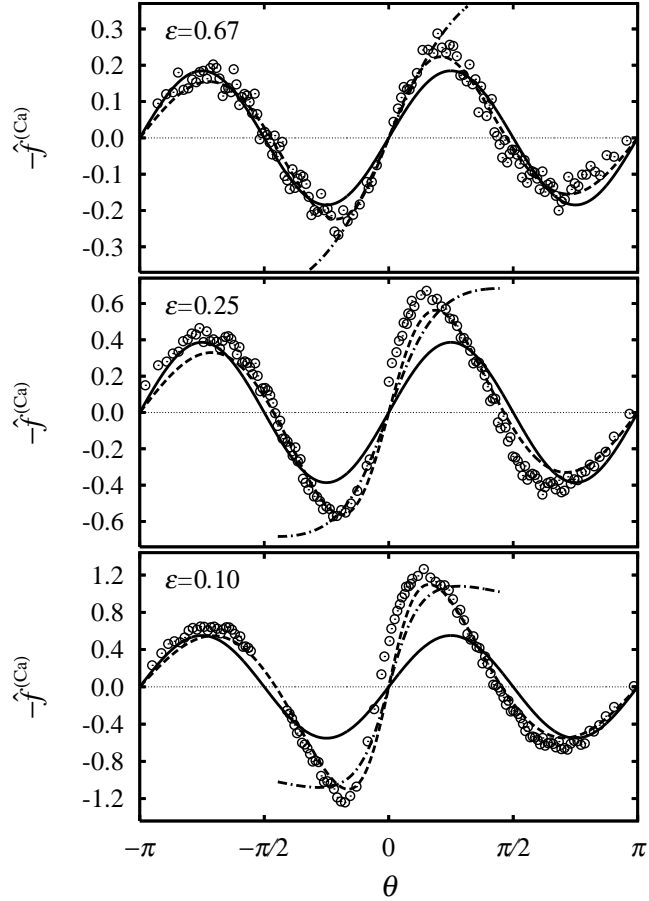,height=9cm,angle=270}

\end{center}
\caption{
Angular profile of the deflection. 
The solid curves indicate the analytical solution (\cite{mag2003}) 
$-\hat{f}^{({\rm Ca})}=$
$\frac{3}{4}\kappa^2$
$\{1+\frac{3}{8}\kappa$
$(1+\frac{3}{8}\kappa+\frac{73}{64}$
$\kappa^2)\}\sin\theta\cos\theta$ 
with the assumption of the sufficiently long distance between the bubble 
 centroid and the wall, i.e., $\kappa (=(1+\varepsilon)^{-1})\ll 1$.
The circles indicate the experimental results (\cite{tak2002}),
 the dashed curves the results obtained by the numerical simulation,
and the dashed-dotted curves the prediction 
(\ref{eq:f_lub})
($-\hat{f}^{({\rm Ca})}=\frac{3}{5\theta}\log(1+\frac{\theta^2}{2\varepsilon})$)
by the lubrication approach (\cite{hod2004}). 
Upper, middle, and lower panels show the results at 
$\varepsilon=0.67$, $\varepsilon=0.25$, and $\varepsilon=0.10$, respectively,
and the corresponding capillary numbers in the experiment are 
${\rm Ca}=0.080$, ${\rm Ca}=0.068$, and ${\rm Ca}=0.056$, respectively.
}
\label{fig:scale_deform}
\end{figure}

\subsection{Interfacial deformation}

To demonstrate the narrow-gap effect, we investigate the bubble deformation.
We here examine the scaled interfacial deflection 
$\hat{f}^{({\rm Ca})}(\theta)=f(\theta,\phi)/({\rm Ca}\ \cos\phi)$.
In the experiment, we estimated $f(\theta,\phi)$ 
taking a circumference of the bubble on the plane $x_2=0$. 
Figure \ref{fig:scale_deform} shows the angular profile of 
the deflection $-\hat{f}^{({\rm Ca})}$ for various $\kappa (=(1+\varepsilon)^{-1})$.
As shown in figure \ref{fig:scale_deform}(a) for the relatively wide gap $\varepsilon=0.67$,
the analytical solution for $\kappa\ll 1$ (\cite{mag2003})
is consistent with the measured and simulated deflections.
Note that the agreement between the theoretical and simulated deflections
is confirmed to be better in the wider separation.
For the narrower gap ($\varepsilon=0.10$, $0.25$), by contrast, 
the analytical solution of the deflection magnitude 
is smaller than the measured one especially in the wall neighborhood ($\theta\sim 0$), 
which may be related to the considerable underestimation of the migration velocity as
shown in figure \ref{fig:migvel_comp_exp}.
Contrastingly, the present simulation quantitatively captures 
the local near-wall profile of the measured deflection as well as the global magnitude. 

\begin{figure}
\begin{center}

\epsfig{file=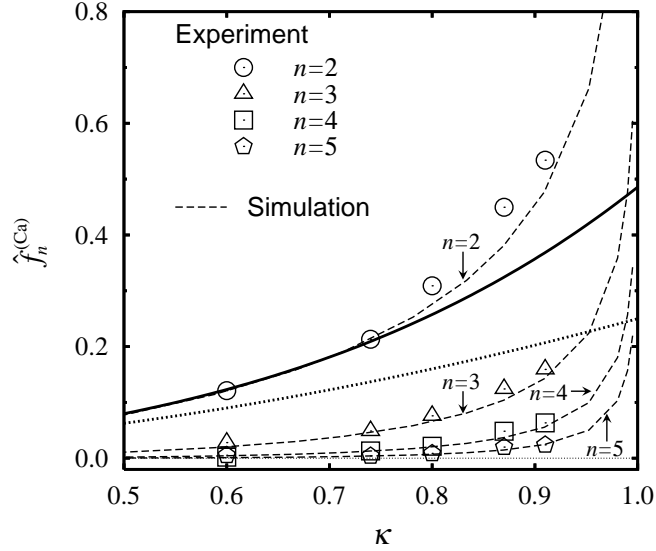,height=9cm,angle=270}

\end{center}
\caption{
Modal deflection $\hat{f}_n^{({\rm Ca})}$,
which is expanded in associated Legendre polynomials $P_n^1$,
versus $\kappa (=(1+\varepsilon)^{-1})$.
The dotted curve indicates the leading term of 
the analytical solution (\cite{mag2003})
$\hat{f}_n^{({\rm Ca})}=\frac{1}{4}\kappa^2$, 
and the solid curve the analytical solution
with the higher-order contributions 
$\hat{f}_n^{({\rm Ca})}=\frac{1}{4}\kappa^2$
$\{1+\frac{3}{8}\kappa$
$(1+\frac{3}{8}\kappa+\frac{73}{64}$
$\kappa^2)\}$.
The symbols indicate the experimental results (\cite{tak2002}),
and the dashed curves the results obtained by the numerical simulation.
}
\label{fig:deformationr_kap}
\end{figure}

To quantify the local effect of such a large discrepancy 
in $\hat{f}^{({\rm Ca})}$ on the migration velocity, 
we here describe the deflection in an expansion form
\begin{equation}
\hat{f}^{({\rm Ca})}(\theta)=
\sum_{n=2}^\infty
\hat{f}_n^{({\rm Ca})}
P_n^1(\cos\theta),
\end{equation}
where $P_n^1$ represents the associated Legendre polynomial.
Figure \ref{fig:deformationr_kap} shows the modal deflections 
$\hat{f}_n^{({\rm Ca})}$ for $2\leq n\leq 5$ as a function of $\kappa$.
The simulation results are consistent with the measured deflections for
all the shown modes.
In predicting the bubble migration, 
the wide-gap theory (\cite{mag2003}) assumes that 
the mirror image primarily induces the deformation of the mode $n=2$.
The leading-order of the $n=2$ deflection is $\hat{f}_2^{({\rm Ca})}=\kappa^2/4$ 
in the limit of $\kappa\rightarrow 0$.
Considering the higher-order effect with respect to $\kappa$, 
\cite{mag2003} derived 
$\hat{f}_n^{({\rm Ca})}=\frac{1}{4}\kappa^2$
$\{1+\frac{3}{8}\kappa$
$(1+\frac{3}{8}\kappa+\frac{73}{64}$
$\kappa^2)\}$.
For $\kappa < \sim 0.7$, such a higher-order $\kappa$ correction
is responsible for the enhancement of the $n=2$ deflection from the leading-order one
as seen in the better agreement with the measured and simulated deflections.
However, 
the correction is not sufficient for the narrower gap $\kappa> \sim 0.7$, 
and thus the theoretical underestimation of the $n=2$ deflection 
becomes more serious with $\kappa$.
Further, the theory does not cover the considerable increase 
in the higher-order $n\geq 3$ deflections with $\kappa$
as demonstrated by both the measurement and the simulation.

The modal deflection $\hat{f}_n^{({\rm Ca})}$
is linked to the migration velocity and force as decomposed into
\begin{equation}
V_{B3}^{({\rm Ca})}=\sum_{n=2}^\infty V_{B3,n}^{({\rm Ca})},\ \ 
F_M^{({\rm Ca})} = \sum_{n=2}^\infty F_{M ,n}^{({\rm Ca})}, 
\end{equation}
which are 
\begin{equation}
\frac{V_{B3,n}^{({\rm Ca})}}{V_{B1}}=\frac{F_{M ,n}^{({\rm Ca})}}{F_{DC}},
\label{eq:vb3nca_fm}
\end{equation}
\begin{equation}
F_{M ,n}^{({\rm Ca})}=
\oint_{S_B}\!\!\!\!\!{\rm d}^2{\bm x}\ 
{\cal L}\left(
\hat{f}_n^{({\rm Ca})}P_n^1(\cos\theta)\cos\phi
\right).
\end{equation}

\begin{figure}
\begin{center}

\epsfig{file=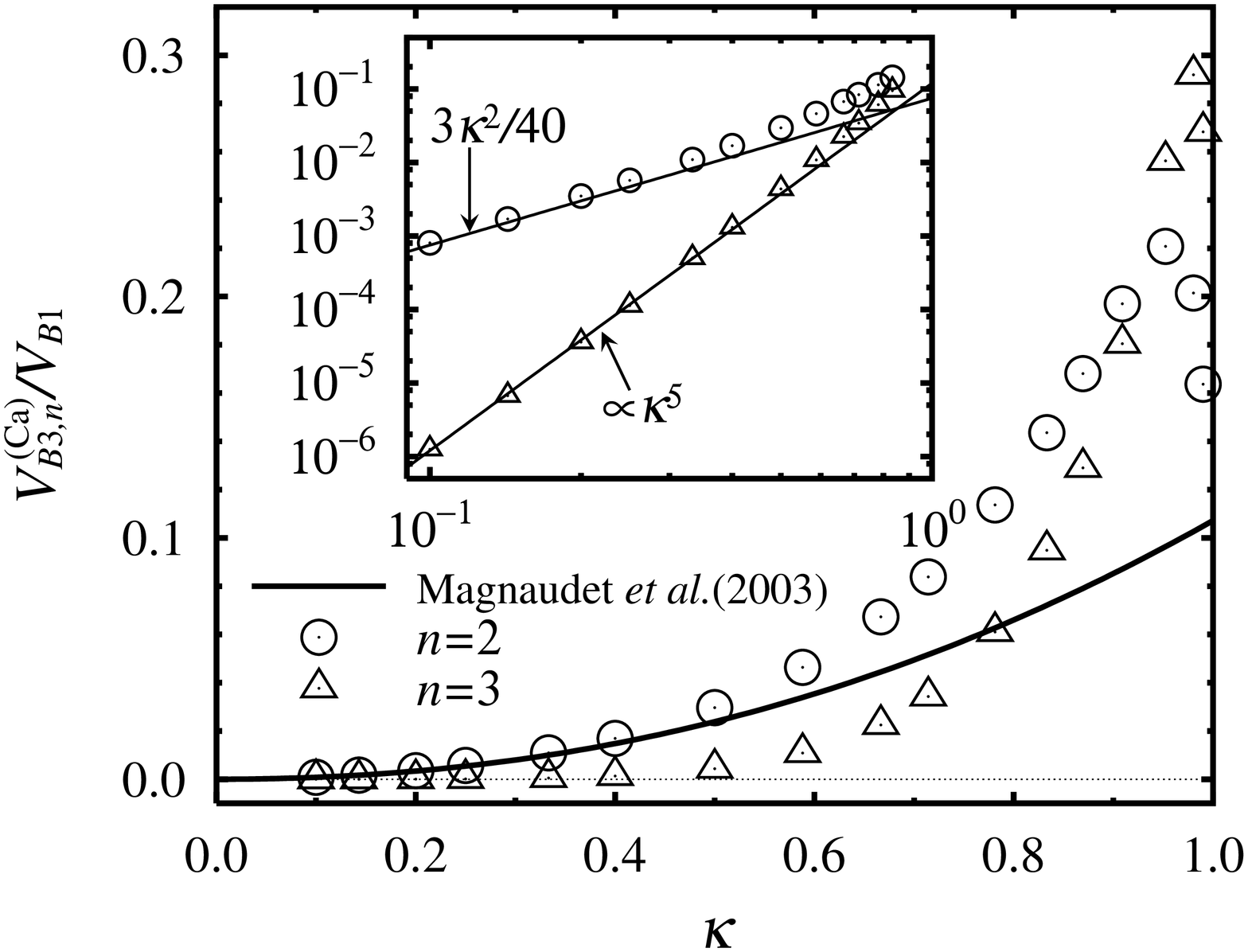,width=7cm,angle=270}

\epsfig{file=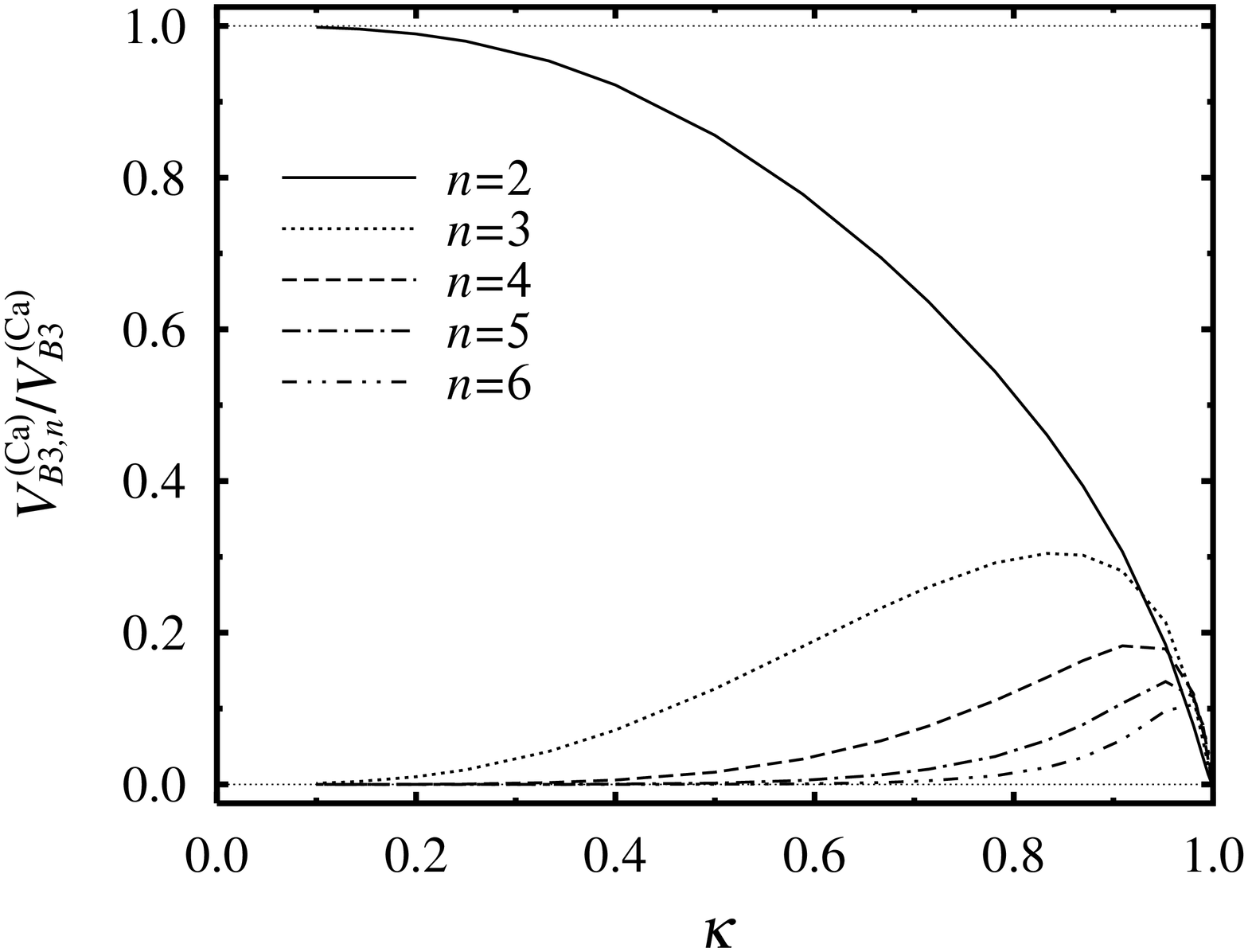,width=7cm,angle=270}
\end{center}
\caption{
Contribution of the modal deflection to the migration velocity
 $V_{B3,n}^{({\rm Ca})}$, 
which is expanded in associated Legendre polynomials $P_n^1$. 
(a) top panel: 
the linear-linear plot of 
the velocity of the modes $n=2$ and $n=3$ versus $\kappa (=(1+\varepsilon)^{-1})$.
The solid line indicates the analytical solution (\cite{mag2003})
$V_{B3,2}^{({\rm Ca})}/V_{B1}=$
${3\kappa^2(1+3\kappa/2)}$
$/\{40(1+3\kappa/4)\}$.
The symbols indicate the results obtained by the numerical simulation.
The inset shows the same plot in a log-log scale.
(b) bottom panel: 
The simulation results of 
the modal contribution normalized by the total velocity.
}
\label{fig:modal}
\end{figure}

\noindent
Figure \ref{fig:modal}(a) shows the contribution of the modal deflection 
to the migration velocity for the modes $n=2$ and $n=3$. 
For small $\kappa$, the simulation result is consistent with the
analytical solution (\cite{mag2003}), 
which considers only the $n=2$ deformation to cause the bubble migration.
The inset shows that for small $\kappa$, the contribution of the mode $n=2$ 
is proportional to $\kappa^2$, 
while that of the mode $n=3$ to $\kappa^5$, 
whose exponent is not trivially proven but may be predictable
extending the regular perturbation to the higher-order. 
The difference in the exponent ensures that 
the relative contribution of the mode $n=3$ to $n=2$
becomes more significant with $\kappa$.
It should be noticed that 
although we confirmed that the migration force contribution $F_{M, 2}^{({\rm Ca})}$ 
of the mode $n=2$ increases
as $\kappa\rightarrow 1$ (i.e., $\varepsilon\rightarrow 0$), 
the velocity contribution $V_{B3, 2}^{({\rm Ca})}$ reduces 
as shown in figure \ref{fig:modal}(a).
It is because the slope of the $n=2$ migration force, 
$-{\rm d}\log F_{M, 2}^{({\rm Ca})}/{\rm d}\log \varepsilon$,
in a logarithmic plot is more gentle than that of the drag force,
$-{\rm d}\log F_{DC}/{\rm d}\log \varepsilon\rightarrow 1$
in the denominator of (\ref{eq:vb3nca_fm})
as $\varepsilon\rightarrow 0$.
Figure \ref{fig:modal}(b) shows the modal contribution $V_{B3,n}^{({\rm Ca})}$
compared with the migration velocity $V_{B3}^{({\rm Ca})}$.
The contribution of the mode $n=2$ monotonically decreases with $\kappa$.
The contribution of the mode $n=3$ increases with $\kappa$ in the range
of $\kappa< \sim 0.8$, while decreases in the greater $\kappa$.
It is because the higher modal contributions for $n\geq 4$ become no longer disregarded. 
Moreover, 
the fact that all the shown modal contributions decay as $\kappa\rightarrow 1$
indicates that 
the further higher-order contributions become considerable, 
and the regular perturbation approach with respect to $\kappa$ is no longer effective.

\subsection{Comparison with small deformation theory in the lubrication limit}
\label{sec:abe_lub}

Figures \ref{fig:scale_deform}, \ref{fig:deformationr_kap} and \ref{fig:modal}
imply that for small $\varepsilon$, 
the bubble deformation is preferentially enhanced within the narrow bubble-wall gap, 
and then its squeezing effect promotes the bubble migration. 
To shed more light on the role of the hydrodynamics in the gap,
 comparisons with a small deformation theory in the lubrication limit will be made. 
It should be noticed that \cite{hod2004} have performed a lubrication study 
 for a nearly spherical drop near a tilted plane in so-called `slipping' regime,
 and derived the deformation-induced normal force.
One can also access a relevant physical picture in theoretical studies on the lift force on an elastic body
 induced by its deformation (\cite{sek1993, sko2004, sko2005, urz2007}).
Following the spirit of the lubrication theory, we evaluate the migration velocity
in the limit of $\varepsilon\rightarrow 0$.

The basic equations for the lubrication analysis
 and the solutions are shown in Appendix \ref{appendix_b}.
To be recalled and to be used for comparison with the simulation results,
 the preconditions and the perturbed quantities are detailed here.
We prescribe the wall-parallel velocity $V_{B1}$,
and employ standard lubrication assumption, i.e., $\varepsilon \ll 1$. 
We also suppose small capillary number ${\rm Ca}\ll 1$. 
As implied in (\ref{eq:def_f_lub}),
 the deflection is $O({\rm Ca}\varepsilon^{-1/2}a)$,
 which has to be sufficiently smaller than the gap $\varepsilon a$
 if the tilt angle of the near-wall interface from the plane wall 
 is supposed to be small. 
Here we adopt an additional constraint $\delta\equiv\varepsilon^{-3/2}{\rm Ca}\ll 1$.
For inner coordinates $(R,Z,\phi)$
(here $\varepsilon^{1/2}R=r$ and $\varepsilon^{-1}Z=z$ as illustrated 
in figure \ref{fig:schem}(b), (see e.g. \cite{gol1967, one1967})),
a parabolic profile $Z=H(R)\equiv 1+R^2/2$ represents the interface 
within the inner region $r\sim \varepsilon^{1/2}$, 
if the deformation is absent. 
Using $\varepsilon$, 
we write the velocity vector, the pressure, and the deflection $f^{\rm (Ca)}$ of the
interface in an expansion form with respect to $\delta$
\begin{equation}
\begin{split}
u_r(r,\phi,z)=&
\hat{U}_r^{(0)}(R,Z)\cos\phi+
\delta\hat{U}_r^{(f)}(R,Z)
\\&
+\varepsilon\hat{U}_r^{(1)}(R,Z)\cos\phi
+\delta\hat{U}_r^{(f,2)}(R,Z)\cos2\phi
+O(\delta^2)+O(\varepsilon\delta)+...,
\label{eq:ur_lub}
\end{split}
\end{equation}
\begin{equation}
\begin{split}
u_\phi(r,\phi,z)=&
\hat{U}_\phi^{(0)}(R,Z)\sin\phi\\
&+\varepsilon\hat{U}_\phi^{(1)}(R,Z)\sin\phi
+\delta\hat{U}_\phi^{(f,2)}(R,Z)\sin2\phi
+O(\delta^2)+O(\varepsilon\delta)+...,
\end{split}
\end{equation}
\begin{equation}
\begin{split}
u_z(r,\phi,z)=&
\varepsilon^{1/2} \left(\hat{U}_z^{(0)}(R,Z)\cos\phi+
\delta\hat{U}_z^{(f)}(R,Z)\right.\\
&\left.+\varepsilon\hat{U}_z^{(1)}(R,Z)\cos\phi
+\delta\hat{U}_z^{(f,2)}(R,Z)\cos2\phi
+O(\delta^2)+O(\varepsilon\delta)+...\right),
\end{split}
\end{equation}
\begin{equation}
\begin{split}
p(r,\phi,z)=&
\varepsilon^{-3/2} \left(\hat{P}^{(0)}(R)\cos\phi
+\delta \hat{P}^{(f)}(R)\right.\\
&\left.+\varepsilon\hat{P}^{(1)}(R,Z)\cos\phi
+\delta\hat{P}^{(f,2)}(R,Z)\cos2\phi
+O(\delta^2)+O(\varepsilon\delta)+...\right),
\label{eq:p0_lub}
\end{split}
\end{equation}
\begin{equation}
f^{\rm (Ca)}(r,\phi)=
\varepsilon^{-1/2}
\left(
\hat{F}(R)\cos\phi
+O(\delta)+O(\varepsilon)
\right),
\label{eq:def_f_lub}
\end{equation}
whose scaling relations are suitable to all equations
 in Appendix \ref{appendix_b}.
It should be noted that the terms with 
 the superscript $(0)$, $(1)$ or $(f,2)$  in (\ref{eq:ur_lub})--(\ref{eq:p0_lub})
 are proportional to $\cos\phi$, $\sin\phi$,
 $\cos 2\phi$ or $\sin 2\phi$, 
 and thus provide no wall-normal force.
Here we make clear the physical domain of validity of the condition $\delta\ll 1$
 when the rising velocity $V_{B1}$,
 which is needed to evaluate ${\rm Ca}(=\mu V_{B1}/\gamma)$, is unknown. 
Following the analysis by \cite{one1967}, 
 we write the drag force $F_D$ acting on the spherical bubble
 translating parallel to the wall
 in a form $F_{D}=6\pi\mu a V_{B1}(A\log \varepsilon+B)$,
 where $A$ and $B$ are independent of $\varepsilon$.
Considering the free-slip boundary condition on the bubble surface,
 we can analytically find $A=-1/5$.
From our numerical data, we approximately estimate $B=0.6$. 
Hence, from the force balance $F_D=4\pi \rho a^3g/3$, for $\varepsilon\ll 1$,
 we evaluate the rising velocity
$$
V_{B1}=
\frac{1}{(-\log\varepsilon+3)}\frac{10\rho a^2 g}{9\mu},
$$
and obtain the following relation for the bubble radius $a$ 
 to satisfy the condition $\delta\ll 1$.
$$
a \ll \sqrt{\frac{9\gamma}{10\rho g}}\varepsilon^{3/4}(-\log \varepsilon+3)^{1/2}.
$$

\begin{figure}
\begin{center}

\epsfig{file=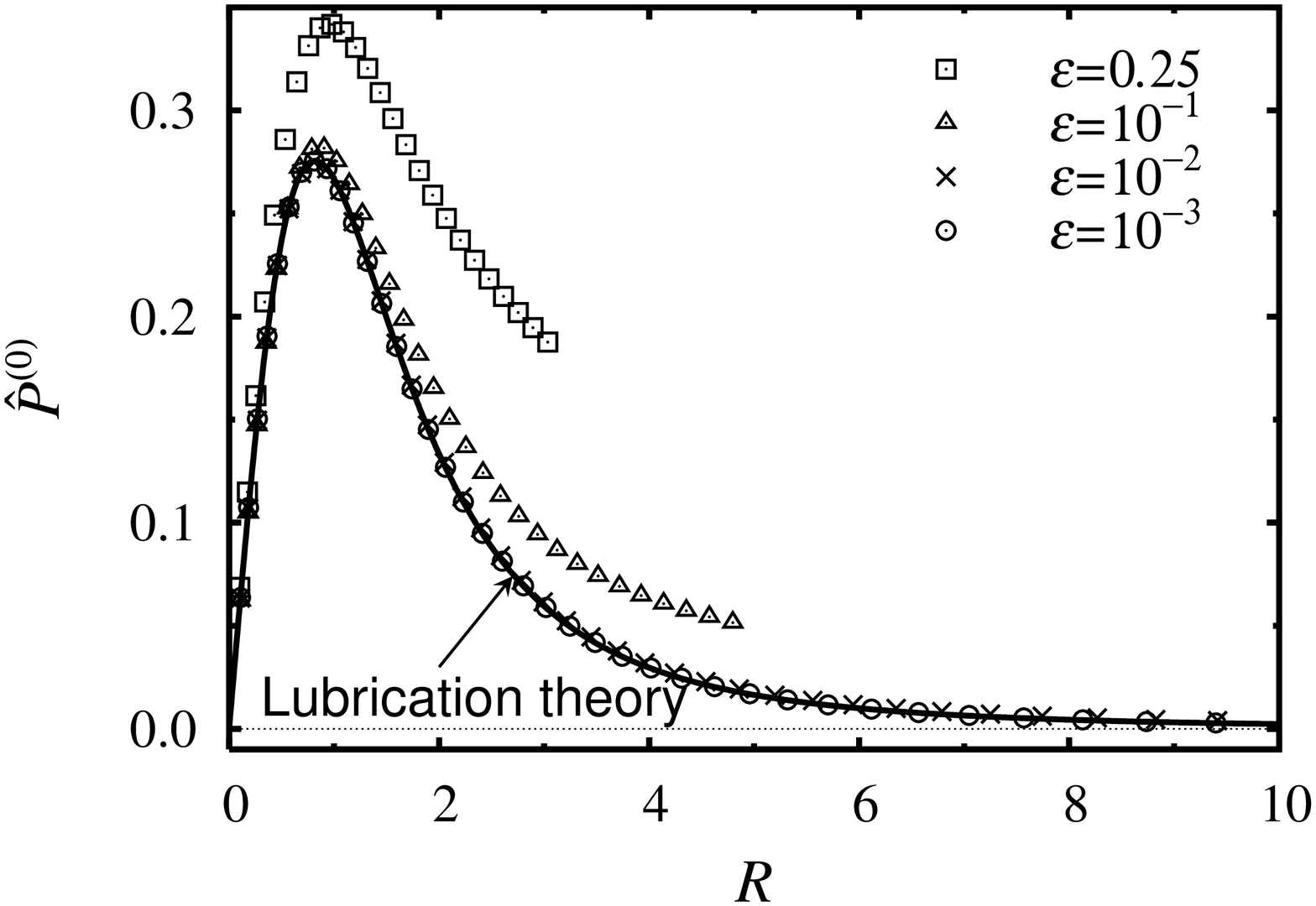,width=7cm,angle=270}

\epsfig{file=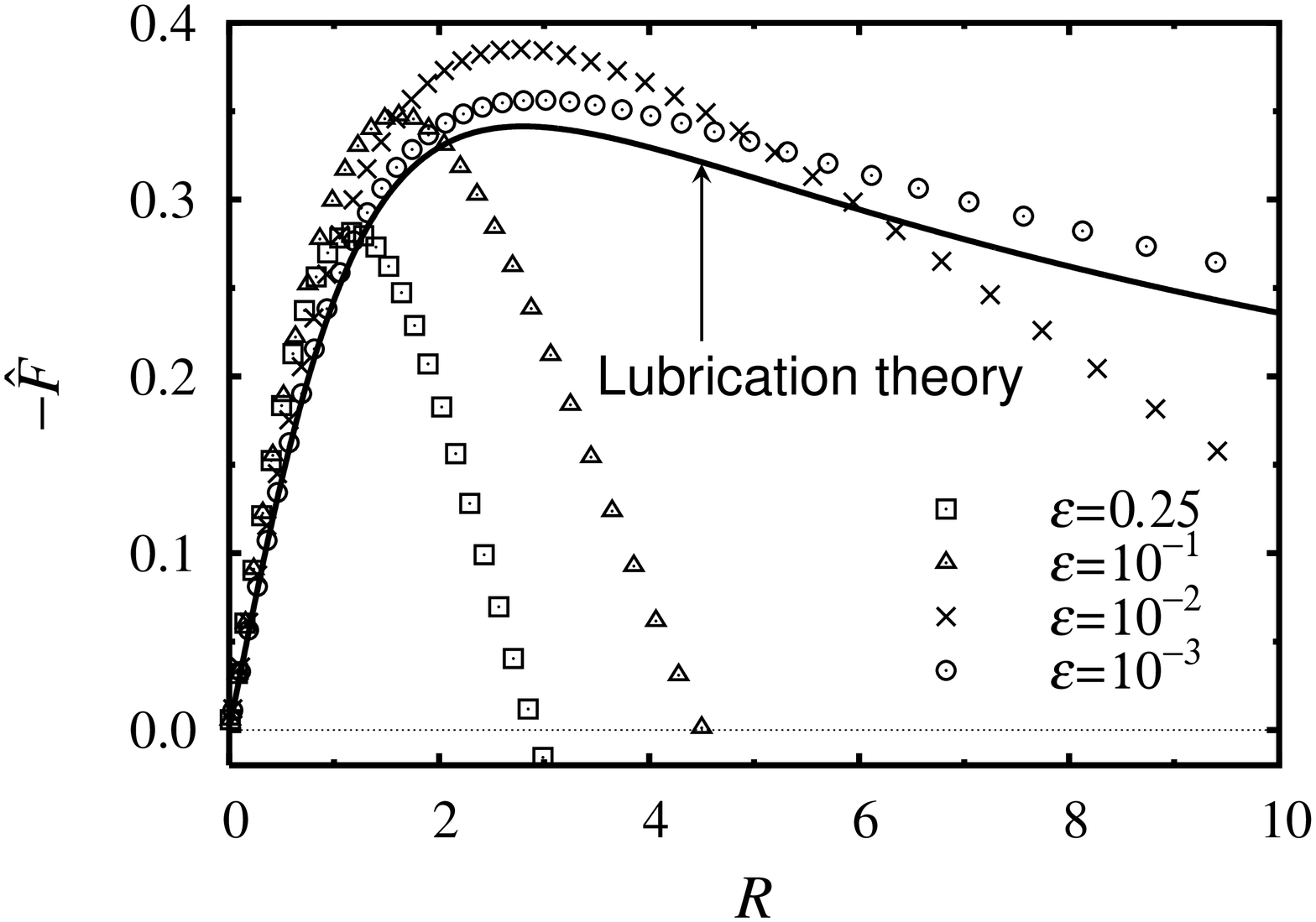,width=7cm,angle=270}

\end{center}
\caption{
Profiles of the pressure ((a) top panel) and the deflection ((b) bottom panel) 
 in the inner expansion scalings of the lubrication theory.
The solid curves indicate the predictions 
(\ref{eq:p_lub}) and (\ref{eq:f_lub}) by the lubrication approach (\cite{hod2004}).
The symbols indicate the results obtained by the numerical simulation.
For the plot, the simulation data of 
the interfacial pressure $P$ and deflection $\hat{f}^{({\rm Ca})}$
for the bubble translating parallel to the wall are scaled as 
$\hat{P}^{(0)}=\varepsilon^{-3/2} P/\cos\phi$ and
$\hat{F}=\varepsilon^{-1/2}\hat{f}^{({\rm Ca})}$,
and the angular coordinate as $R=\varepsilon^{-1/2}\theta$.
}
\label{fig:scale_pres}
\end{figure}

To assure us of the appearance of the lubrication effect, 
figure \ref{fig:scale_pres} shows the profiles of
the pressure $\hat{P}^{(0)}(=3R/(5H^2))$ and deflection $-\hat{F}(=3\log H/(5R))$
in the lubrication limit (see (\ref{eq:p_lub}) and (\ref{eq:f_lub}), respectively),
which are compared with the simulation results of the scaled interfacial pressure
$\varepsilon^{3/2} P/\cos\phi$ and the scaled deflection $\varepsilon^{1/2}\hat{f}^{({\rm Ca})}$
near the wall as a function of $\varepsilon^{-1/2}\theta$. 
For sufficiently small $\varepsilon$, 
the simulation data of the scaled pressure collapse onto 
the curve (\ref{eq:p_lub}) as shown in figure \ref{fig:scale_pres}(a).
Therefore, the scaled deflection profile 
reasonably approaches the lubrication solution (\ref{eq:f_lub})
with decreasing the bubble-wall gap 
as shown in figure \ref{fig:scale_pres}(b).
As plotted as the dashed-dotted curve in figure \ref{fig:scale_deform}, 
the deflection based on (\ref{eq:def_f_lub}) and (\ref{eq:f_lub})
is consistent with the measured and simulated deflections in the wall neighbour ($\theta\sim 0$).
Hence, the discrepancy between the deflections of
 the narrow-gap experiment and the wide-gap theory
 (\cite{mag2003}) is attributable to the lubrication effect.

From \cite{bar1968}, 
the drag force acting on the bubble translating perpendicular to the plane wall is
$F_{DC}\rightarrow 3\pi \mu /(2\varepsilon)$ as
$\varepsilon\rightarrow 0$. 
Substituting this relation and (\ref{eq:fm_lub})
 (i.e., $F_M$ in the lubrication limit) into (\ref{eq:vb3ca}), 
 we obtain the asymptotic solution of the migration velocity
\begin{equation}
\frac{V_{B3}}{V_{B1}}=\frac{F_M}{F_{DC}}
\rightarrow 
\frac{{\rm Ca}}{25}\varepsilon^{-1}
\ \ \ {\rm as}\ \ \ 
\varepsilon\rightarrow 0.
\label{eq:vb3_lub}
\end{equation}

\begin{figure}
\begin{center}

\epsfig{file=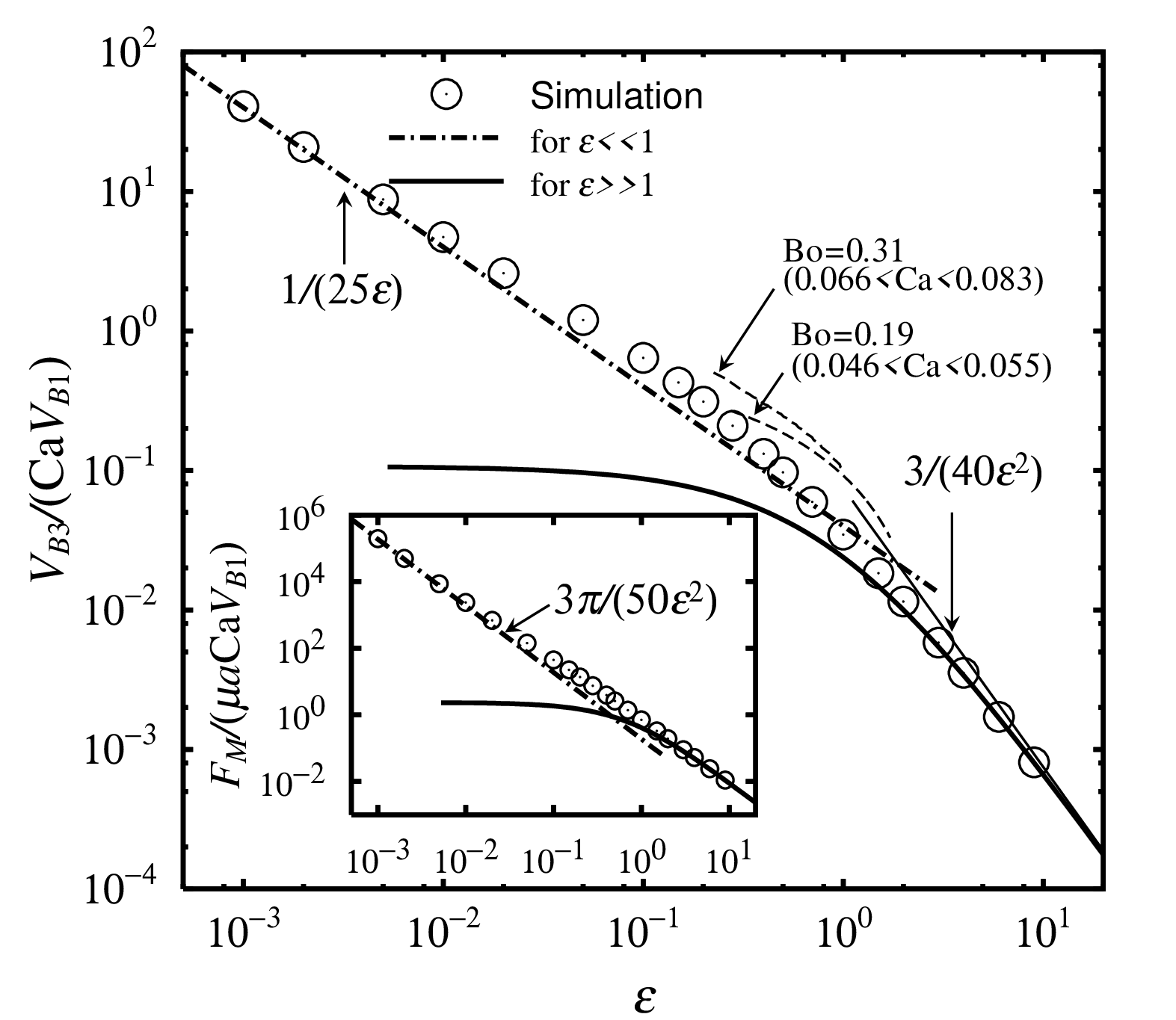,width=9cm,angle=270}

\end{center}
\caption{
The lateral migration velocity versus the clearance parameter $\varepsilon$. 
The circles indicate the results obtained by the numerical simulation.
The solid, and dashed-dotted curves correspond to the analytical solution 
$V_{B3}=3{\rm Ca}V_{B1}\kappa^2$
$(1+3\kappa/2)/\{40(1+3\kappa/4)\}$ (\cite{mag2003}), 
and 
$V_{B3}=\frac{1}{25}{\rm Ca}V_{B1}\varepsilon^{-1}$
derived by means of the lubrication approach (\cite{hod2004}), respectively. 
The dashed curves indicate the experimental results (\cite{tak2002}).
The inset shows the lateral force (\ref{eq:fm_lub}). 
}
\label{fig:migvel_comp_lub}
\end{figure}

\noindent
To make comparisons with the asymptotic solutions,
 figures \ref{fig:migvel_comp_exp} and \ref{fig:migvel_comp_lub} 
 show the scaled migration velocity
 as functions of the inverse distance $\kappa (=(1+\varepsilon)^{-1})$ and 
 the clearance parameter $\varepsilon$, respectively.
The inset of figure \ref{fig:migvel_comp_lub} shows the scaled force (\ref{eq:fm_lub}). 
The simulation results are consistent with 
two asymptotic behaviors based on the lubrication theory
for $\varepsilon\ll 1$  as well as the mirror image technique 
for $\varepsilon \gg 1$ (i.e., $\kappa\ll 1$). 
As seen from figure \ref{fig:migvel_comp_lub}, 
 the theories provide the different exponents of the migration velocity
 scaling with respect to $\varepsilon$, namely,
$V_{B3}/V_{B1}\propto {\rm Ca}\ \varepsilon ^{-2}$ for $\varepsilon\gg 1$
and $V_{B3}/V_{B1}\propto {\rm Ca}\ \varepsilon ^{-1}$ for $\varepsilon\ll 1$. 

The puzzling finding in \cite{tak2009} that 
the theory of \cite{mag2003} accurately predicts the deformation
but fails to predict quantitatively the deformation-induced migration velocity
is explained from the fact that 
the ratio of the simulation result of $V_{B3}$ to the theoretical prediction
is more sensitively dependent upon $\varepsilon$ than that of
$\hat{f}_2^{({\rm Ca})}$ when the lubrication effect becomes relevant. 
For instance, for $\varepsilon=0.4$, $\varepsilon=0.2$ and $\varepsilon=0.1$, 
the simulation-to-theory ratios of $\hat{f}_2^{({\rm Ca})}$ are respectively $1.004$, $1.1$ and $1.3$ (figure \ref{fig:deformationr_kap}), 
while those of $V_{B3}$ are respectively $2.5$, $4.3$ and $7.3$ (figure \ref{fig:migvel_comp_exp}).
The lubrication effect is likely to compensate
 the large discrepancy between the migration 
 velocities of the experiment and the wide-gap theory revealed in \cite{tak2009}. 
However, although the quantitative agreement between the interfacial deflections
of the experiment and the simulation is shown in figures \ref{fig:scale_deform} and \ref{fig:deformationr_kap},
the migration velocity in the experiment is still considerably higher than the simulated one. 
Its cause is not clear at this moment, 
 and further joint researches among theory, numerics and experiment are
 needed to resolve this discrepancy problem.

\section{Conclusion}

We numerically and theoretically investigated 
deformation-induced lateral migration of a bubble slowly rising near a vertical
plane wall in a stagnant liquid. 
We focused on a situation with a short clearance $c$
between the bubble interface and the wall.
We demonstrated that the 
 wide-gap theory (\cite{mag2003}),
 which considers the $n=2$ deformation mode,
 describes the deformation-induced lift force as long as 
 the bubble-wall gap is sufficiently wide
 ($a/(a+c)\ll 1$, here $a$ is the bubble radius). 
For the narrow-gap case with the clearance parameter $\varepsilon(= c/a)$ smaller than unity, 
 we found that the higher-order $n\geq 3$ deformation modes
 crucially enhance the migration velocity,
 and the lubrication effect (\cite{hod2004})
 appears to induce the migration velocity, which scales asymptotically like
 $V_{B3}\rightarrow {\rm Ca}\ \varepsilon^{-1}V_{B1}/25$
 as $\varepsilon\rightarrow 0$.
This contrasts with the case of the inertia-driven migration,
 to which the wide-gap theory demonstrated a robust applicability
 in prediction over a wide range of $\varepsilon$.

The present simulation consistently served as bridge between the wide- and narrow-gap theories 
(see figure \ref{fig:migvel_comp_lub})
as long as the bubble deformation is assumed to be infinitesimal.
However, in spite of the qualitative success of the simulation
in revealing the narrow-gap repulsive force, 
the deformation-induced migration velocity in the experiment
is considerably higher by a factor of about 3 than the simulated one, as shown in figure \ref{fig:migvel_comp_exp}.
The experiment may inevitably involve unknown factors
such as unsteadiness, imperfection from the infinite plate-fluid system, and measurement uncertainty,
which cannot be captured by the simulation. 
Nevertheless, we have not expected such a large discrepancy
because firstly the quantitative agreement between the interfacial deflections
of the experiment and the simulation has been confirmed
 in figures \ref{fig:scale_deform} and \ref{fig:deformationr_kap}, 
secondly the consistent inertia-driven migration velocity of a rigid sphere with the available theories
(\cite{vas1976, mag2003})
has been obtained by \cite{tak2004} using the same experimental setup,
and thirdly considerable uncertainty seems not to be introduced
into such a simple system as illustrated in figure \ref{fig:schem}.
Further joint researches among theory, numerics and experiment are
needed to resolve this problem. 
As a possible factor causing the inconsistency, 
we raise a difference between the bubble deformation levels of the experiment and the present analysis.
As stated in \S \ref{sec:abe_lub}, 
the infinitesimal deformation assumption in the lubrication limit
is justified only for the case that $\delta(=\varepsilon^{-3/2}{\rm Ca})\ll 1$.
Beyond this limitation, unexplored hydrodynamic ingredients 
possibly become important on the bubble migration.
For the experimental data shown in figure \ref{fig:migvel_comp_exp}
 and figure \ref{fig:migvel_comp_lub}, 
 the maximum value of $\delta$ is $0.74$,
 which is less than but comparable to unity. 
Therefore, 
the bubble deformation is finite rather than infinitesimal,
and is likely to induce the higher-order force, 
which is possibly comparable to or stronger than the leading migration force
evaluated with the infinitesimal deformation theories.
From the theoretical viewpoint, 
a tiny bubble experiment, which results in a tiny capillary number
and thus a tiny deformation, is favorable for comparative study. 
However, such an experiment has often resulted in an undetectably low migration velocity
and made an accurate measurement difficult.
To overcome such a dilemma, the highly accurate boundary element computations 
(e.g. \cite{wan2006, dim2007}) 
for various deformation levels would be helpful to complement the infinitesimal deformation theories.

\begin{acknowledgments}
\noindent
We thank Shu Takagi for fruitful discussion. 
\end{acknowledgments}

\begin{appendix}
\section{Finite-difference descriptions of the basic equation set in bipolar coordinates}
\label{appendix_a}

\begin{figure}
\begin{center}

\vspace{1em}

\epsfig{file=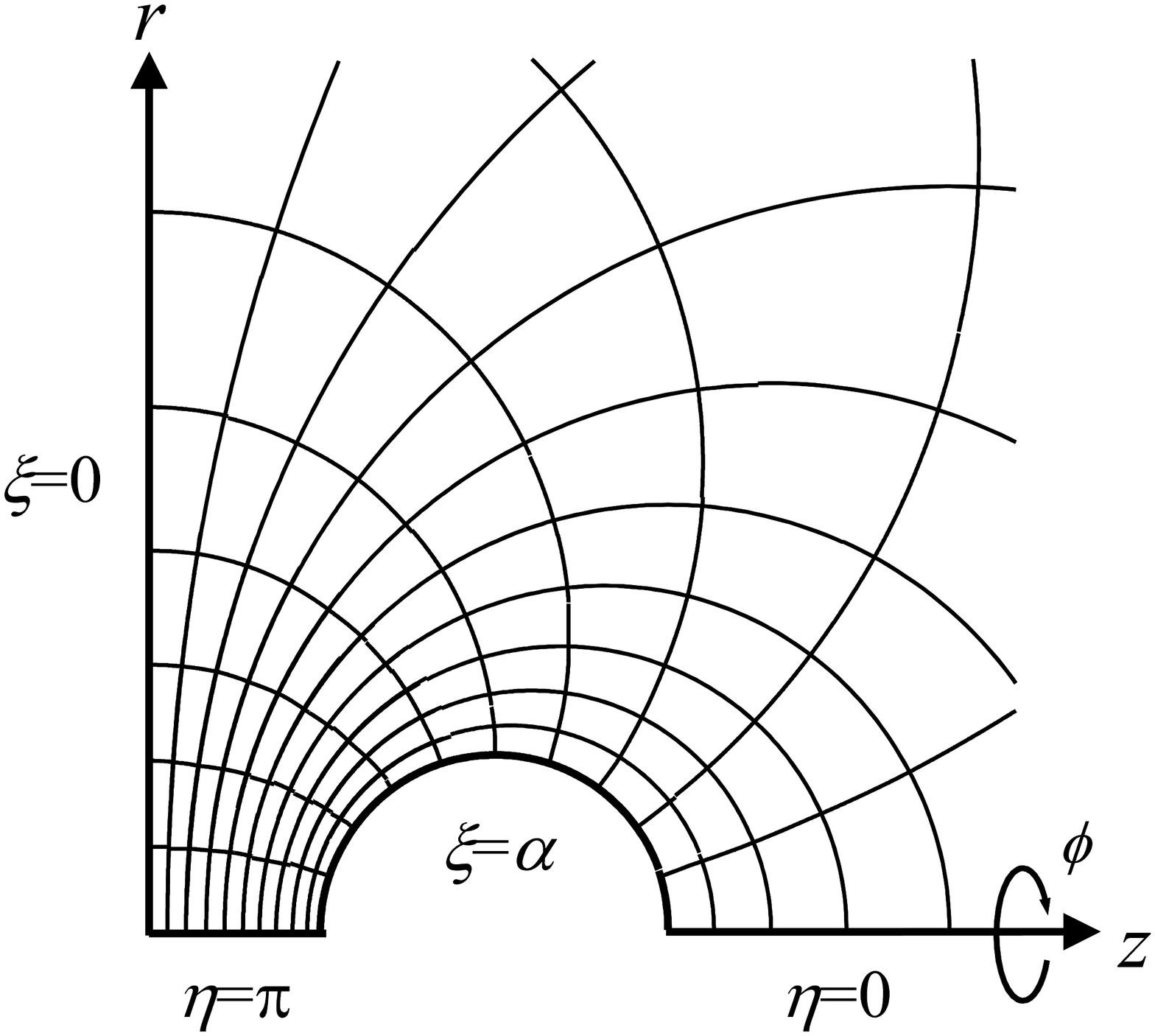,width=6cm,angle=0}
\ \ \ 
\epsfig{file=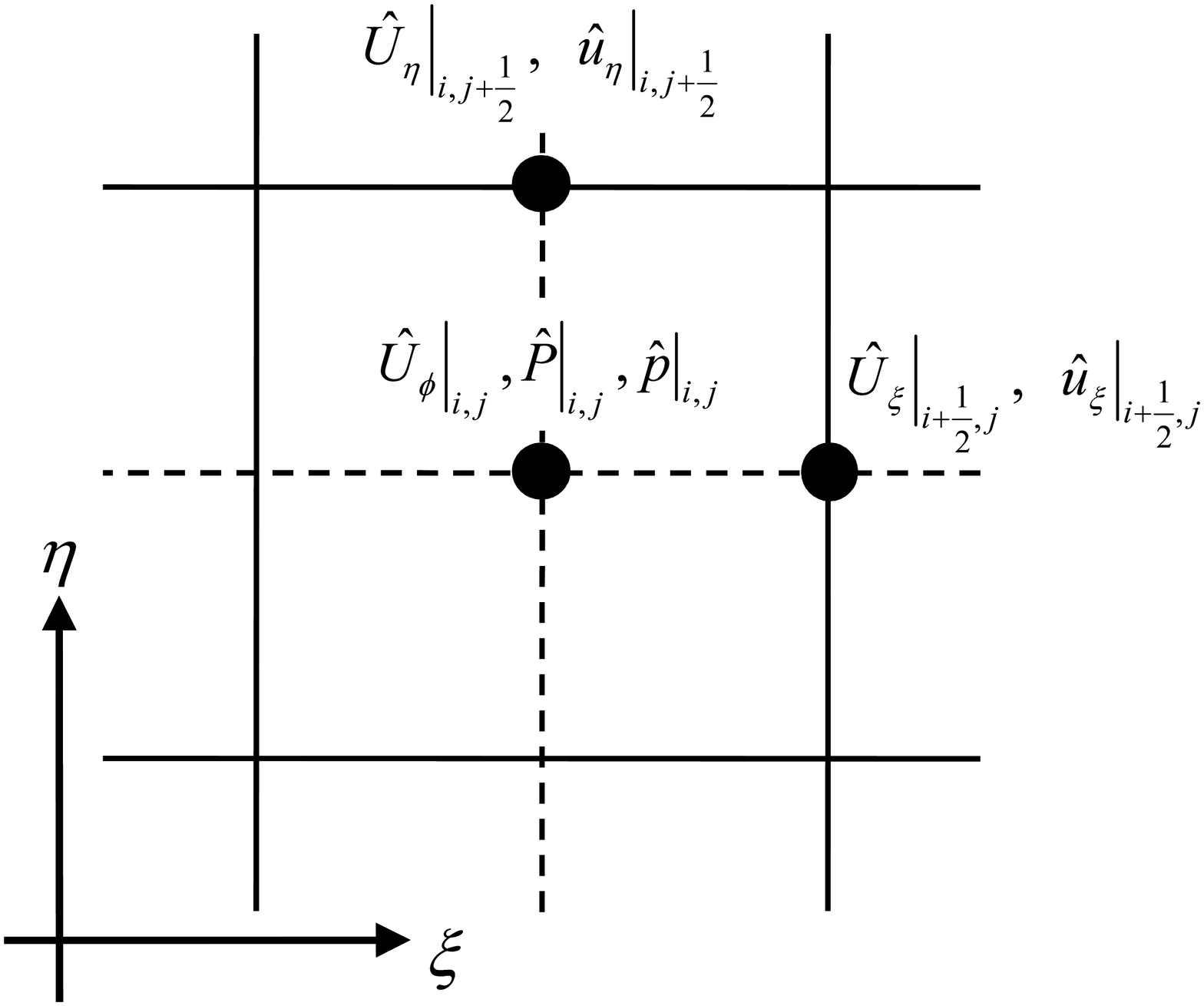,width=6cm,angle=0}
\end{center}
\caption{
Computational grid. 
(a) left panel: The bipolar coordinates.
$\xi=0$ and $\xi=\alpha$ represent 
the plane wall and the bubble surface, respectively.
(b) right panel: 
Definition points of the velocity components and the pressures 
on the staggered grid in a computational space.
}
\label{fig:schem_bipolar}
\end{figure}

We describe the basic equation set in the bipolar coordinates 
(see e.g. \cite{one1964, hap1973}\ Appendix A-19) as illustrated in figure \ref{fig:schem_bipolar}(a).
The coordinates $(r,z)$ in figure \ref{fig:schem} (b) are 
\begin{equation}
r=\frac{k\sin\eta}{{\cal D}},\ \ 
z=\frac{k\sinh\xi}{{\cal D}},
\end{equation}
where ${\cal D}=\cosh\xi-\cos\eta$ and $k=\sqrt{\varepsilon(\varepsilon+2)}$. 
The bubble surface is located at $\xi=\alpha\equiv\log(1+\varepsilon+k)$ 
as shown in figure \ref{fig:schem_bipolar}(a).
The gradient of a scalar function $q$ is written as
\begin{equation}
\nabla q=
\frac{{\bm e}_\xi}{h_\xi}\frac{\partial q}{\partial\xi}+
\frac{{\bm e}_\eta}{h_\eta}\frac{\partial q}{\partial\eta}+
\frac{{\bm e}_\phi}{h_\phi}\frac{\partial q}{\partial\phi},
\end{equation}
where 
${\bm e}$ represents the unit vector, and its subscript the corresponding component.
$h$ denotes the scale factor, defined by e.g. 
$h_\xi=\{(\partial x_1/\partial_\xi)^2$ $+(\partial x_2/\partial_\xi)^2$
$+(\partial x_3/\partial_\xi)^2\}^{1/2}$
(see e.g. \cite{bat1967}, Appendix 2). 
Each component is explicitly given by 
\begin{equation}
h_\xi=h_\eta=\frac{k}{{\cal D}}, \ \ \ h_\phi=r.
\end{equation}
The vector and pressure field $({\bm U},P)$ for the bubble translating parallel to
the wall is written using the quantities with hat in a form (e.g. \cite{sug2008})
\begin{equation}
{\bm U}=
\left\{
\left(
{\bm e}_\xi\hat{U}_\xi+{\bm e}_\eta\hat{U}_\eta
\right)\cos\phi+
{\bm e}_\phi\hat{U}_\phi
\sin\phi
-{\bm e}_1
\right\},\ \ \ 
P=\hat{P}\cos\phi,
\end{equation}
of which the Fourier expansion reduces the three-dimensional problem to the two-dimensional one. 
The vector and pressure field $({\bm u},p)$ for the bubble translating
perpendicular to the wall is
\begin{equation}
{\bm u}=
{\bm e}_\xi\hat{u}_\xi+{\bm e}_\eta\hat{u}_\eta+{\bm e}_3,\ \ \ 
p=\hat{p}.
\end{equation}
It should be noticed that 
as opposed to the reference frames in \S \ref{sec:gf},
we take those for 
$(\hat{U}_\xi, \hat{U}_\eta, \hat{U}_\phi)$ and $(\hat{u}_\xi, \hat{u}_\eta)$
viewed from the plane wall for convenience of the simulations.
Consequently, 
$\hat{U}_\xi=\hat{U}_\eta=\hat{U}_\phi=\hat{u}_\xi=\hat{u}_\eta=0$
on the plane wall and on the boundary sufficiently far from the bubble.

We follow a conventional staggered grid arrangement (\cite{har1965}), 
where the velocity component is located on the corresponding cell
interface, and the pressure at the cell centre, 
as shown in figure \ref{fig:schem_bipolar}(b).
The basic equations are discretized by the second-order finite-difference scheme. 
To numerically guarantee the mass and momentum conservations,
and to accurately conduct the numerical integration in computing the drag and migration forces,
we use the exact values of the grid width, 
the cell interfacial area, 
and the control volume in the bipolar coordinates. 
To this end, for the integral of the scale factors
$$
g_\xi(\xi,\eta)=
\int_{\alpha}^\xi\!\!\!{\rm d}\bar{\xi}\ h_\xi(\bar{\xi},\eta),\ \ \ 
g_{\xi\phi}(\xi,\eta)=
\int_{\alpha}^\xi\!\!\!{\rm d}\bar{\xi}\ h_\xi(\bar{\xi},\eta)h_\phi(\bar{\xi},\eta),
$$
$$
g_\eta(\xi,\eta)=
\int_{\pi}^\eta\!\!\!{\rm d}\bar{\eta}\ h_\eta(\xi,\bar{\eta}),\ \ \ 
g_{\eta\phi}(\xi,\eta)=
\int_{\pi}^\eta\!\!\!{\rm d}\bar{\eta}\ h_\eta(\xi,\bar{\eta})h_\phi(\xi,\bar{\eta}),
$$
$$
g_{\xi\eta}(\xi,\eta)=
\int_{\alpha}^\xi\!\!\!{\rm d}\bar{\xi}
\int_{\pi}^\eta\!\!\!{\rm d}\bar{\eta}\
h_\xi(\bar{\xi},\bar{\eta})h_\eta(\bar{\xi},\bar{\eta}),\ \ 
g(\xi,\eta)=
\int_{\alpha}^\xi\!\!\!{\rm d}\bar{\xi}
\int_{\pi}^\eta\!\!\!{\rm d}\bar{\eta}\ h(\bar{\xi},\bar{\eta}),
$$
we use the exact expressions
\begin{equation}
g_\xi=\frac{2k}{\sin\eta}\biggl(
\tan^{-1}\left(\frac{{\cal D}+{\cal C}}{\cal S}\right)
-\tan^{-1}\left(\frac{{\cal D}_\alpha+{\cal C}_\alpha}{\cal S_\alpha}\right)
\biggr),
\end{equation}
\begin{equation}
g_\eta=-\frac{2k}{\sinh\xi}
\tan^{-1}\left(\frac{{\cal D}+{\cal C}}{\cal S}\right),
\end{equation}
\begin{equation}
\begin{split}
g_{\xi\eta}=&k^2\biggl\{
\left(\frac{1}{\sinh^2\xi}-\frac{1}{\sin^2\eta}\right)\tan^{-1}\left(\frac{{\cal D}+{\cal C}}{\cal S}\right)
\\&
-
\left(\frac{1}{\sinh^2\alpha}-\frac{1}{\sin^2\eta}\right)\tan^{-1}\left(\frac{{\cal D}_\alpha+{\cal C}_\alpha}{\cal S_\alpha}\right)
-\left(
\frac{{\cal C}+2}{2{\cal S}}-\frac{{\cal C}_\alpha+2}{2{\cal S}_\alpha}
\right)\biggr\},
\end{split}
\end{equation}
\begin{equation}
g_{\eta\phi}=k^2\left(
-\frac{1}{{\cal D}}+\frac{1}{\cosh\xi+1}
\right),
\end{equation}
\begin{equation}
g_{\xi\phi}=k^2\left\{
\frac{2\cos\eta}{\sin^2\eta}
\left(
\tan^{-1}\left(\frac{{\cal D}+{\cal C}}{\cal S}\right)-\tan^{-1}\left(\frac{{\cal D}_\alpha+{\cal C}_\alpha}{\cal S_\alpha}\right)
\right)
+\frac{\sinh\xi}{{\cal D}\sin\eta}-\frac{\sinh\alpha}{{\cal D}_\alpha\sin\eta}
\right\},
\end{equation}
\begin{equation}
\begin{split}
g=&-\frac{k^3}{\sin^3\eta}\left\{
\left(
\tan^{-1}\left(\frac{{\cal D}+{\cal C}}{\cal S}\right)-\tan^{-1}\left(\frac{{\cal D}_\alpha+{\cal C}_\alpha}{\cal S_\alpha}\right)
\right)\cos\eta+\frac{{\cal S}}{2{\cal D}}-\frac{{\cal S}_\alpha}{2{\cal D}_\alpha}
\right\}
\\&
-\frac{k^3}{12}\left(
\tanh^3\frac{\xi}{2}-\tanh^3\frac{\alpha}{2}
\right)
+\frac{k^3}{4}\left(
\tanh\frac{\xi}{2}-\tanh\frac{\alpha}{2}
\right),
\end{split}
\end{equation}
where 
\begin{equation}
\begin{split}
&h=h_\xi h_\eta h_\phi,\ \ \ 
{\cal C}=\cosh\xi\cos\eta-1,\ \ \ {\cal S}=\sinh\xi\sin\eta,
\\&
{\cal D}_\alpha=\cosh\alpha-\cos\eta,\ \ \ 
{\cal C}_\alpha=\cosh\alpha\cos\eta-1,\ \ \ {\cal S}_\alpha=\sinh\alpha\sin\eta.
\end{split}\end{equation}
We introduce the finite-difference operators 
$\delta_i$ and $\delta_j$, of which the indices $i$ and $j$ correspond
to discretized coordinates along the respective directions $\xi$ and $\eta$,
such as
\begin{equation}
\left.
\begin{array}{rcl}
\delta_i(q)|_{i,j}&=&q_{i+\frac{1}{2},j}-q_{i-\frac{1}{2},j},
\\
\delta_j(q)|_{i,j}&=&q_{i,j+\frac{1}{2}}-q_{i,j-\frac{1}{2}},
\\
\delta_i\delta_j(q)|_{i,j}&=&
q_{i+\frac{1}{2},j+\frac{1}{2}}-
q_{i+\frac{1}{2},j-\frac{1}{2}}-
q_{i-\frac{1}{2},j+\frac{1}{2}}+
q_{i-\frac{1}{2},j-\frac{1}{2}}.
\end{array}
\right\}
\end{equation}
Using these operators, 
we write the divergence of the velocity vector ${\bm U}$ 
in (\ref{eq:cont_st}) as
\begin{equation}
\left.\widehat{\nabla\cdot{\bm U}}\right|_{i,j}
\left(\equiv \frac{\nabla\cdot{\bm U}}{\cos\phi}\right)
=
\frac{\left(
\delta_i(\delta_j(g_{\eta\phi})\hat{U}_\xi)|_{i,j}
+
\delta_j(\delta_i(g_{\xi\phi})\hat{U}_\eta)|_{i,j}
+
\hat{U}_\phi\delta_i\delta_j(g_{\xi\eta})|_{i,j}
\right)}{\delta_i\delta_j(g)|_{i,j}},
\end{equation}
and the components of the Stokes equation (\ref{eq:mom_st}) for the bubble translating parallel to the
plane wall as
\begin{equation}
\left.
\begin{array}{rcl}
0&=&\displaystyle
\frac{\delta_i(-\hat{P}+\widehat{\nabla\cdot{\bm U}})|_{i+\frac{1}{2},j}}{\delta_i(g_\xi)|_{i+\frac{1}{2},j}}
+\frac{-\delta_j(r\hat{\Omega}_\phi)|_{i+\frac{1}{2},j}+\hat{\Omega}_\eta\delta_j(g_\eta)|_{i+\frac{1}{2},j}}{\delta_j(g_{\eta\phi})|_{i+\frac{1}{2},j}},
\\
0&=&\displaystyle
\frac{\delta_j(-\hat{P}+\widehat{\nabla\cdot{\bm U}})|_{i,j+\frac{1}{2}}}{\delta_j(g_\eta)|_{i,j+\frac{1}{2}}}
+\frac{\delta_i(r\hat{\Omega}_\phi)|_{i,j+\frac{1}{2}}-\hat{\Omega}_\xi\delta_i(g_\xi)|_{i,j+\frac{1}{2}}}{\delta_i(g_{\xi\phi})|_{i,j+\frac{1}{2}}},
\\
0&=&\displaystyle
\frac{(\hat{P}-\widehat{\nabla\cdot{\bm U}})|_{i,j}}{r|_{i,j}}
+\frac{-\delta_i(\delta_j(g_\eta)\hat{\Omega}_\eta)|_{i,j}+\delta_j(\delta_i(g_\xi)\hat{\Omega}_\xi)|_{i,j}}
{\delta_i\delta_j(g_{\xi\eta})|_{i,j}},
\end{array}
\right\}
\end{equation}
where $\hat{\Omega}$ denotes the vorticity, of which each component is 
\begin{equation}
\left.
\begin{array}{rcl}
\hat{\Omega}_\xi|_{i,j+\frac{1}{2}}&=&\displaystyle
\frac{\delta_j(r\hat{U}_\phi)|_{i,j+\frac{1}{2}}+\hat{U}_\eta\delta_j(g_\eta)|_{i,j+\frac{1}{2}}}
{\delta_j(g_{\eta\phi})|_{i,j+\frac{1}{2}}},
\\
\hat{\Omega}_\eta|_{i+\frac{1}{2},j}&=&\displaystyle
\frac{-\delta_i(r\hat{U}_\phi)|_{i+\frac{1}{2},j}-\hat{U}_\xi\delta_i(g_\xi)|_{i+\frac{1}{2},j}}
{\delta_i(g_{\xi\phi})|_{i+\frac{1}{2},j}},
\\
\hat{\Omega}_\phi|_{i+\frac{1}{2},j+\frac{1}{2}}&=&\displaystyle
\frac{\delta_i(\delta_j(g_\eta)\hat{U}_\eta)|_{i+\frac{1}{2},j+\frac{1}{2}}-
\delta_j(\delta_i(g_\xi)\hat{U}_\xi)|_{i+\frac{1}{2},j+\frac{1}{2}}}
{\delta_i\delta_j(g_{\xi\eta})|_{i+\frac{1}{2},j+\frac{1}{2}}}.
\end{array}
\right\}
\end{equation}
Replacing 
$\hat{U}_\xi$, $\hat{U}_\eta$ and $\hat{P}$ 
respectively by
$\hat{u}_\xi$, $\hat{u}_\eta$ and $p$ with 
$\hat{U}_\phi=\hat{\Omega}_\xi=\hat{\Omega}_\eta=0$,
we readily obtain the governing equation 
for the field $(\hat{u}_\xi,\hat{u}_\eta,\hat{p})$.
We write the kinematic and free-slip conditions (\ref{eq:bc_bubsurf_st}) as
\begin{equation}
\hat{U}_\xi|_{N_\alpha+\frac{1}{2},j}=
\widehat{{\bm e}_r\cdot{\bm e}_\xi}|_{N_\alpha+\frac{1}{2},j},
\ \ 
\hat{u}_\xi|_{N_\alpha+\frac{1}{2},j}=
-{\bm e}_z\cdot{\bm e}_\xi|_{N_\alpha+\frac{1}{2},j},
\end{equation}
\begin{equation}
\hat{\Sigma}_{\xi\eta}|_{N_\alpha+\frac{1}{2},j+\frac{1}{2}}
=\hat{\Sigma}_{\xi\phi}|_{N_\alpha+\frac{1}{2},j}
=\hat{\sigma}_{\xi\eta}|_{N_\alpha+\frac{1}{2},j+\frac{1}{2}}=0,
\end{equation}
where the index $N_\alpha+\frac{1}{2}$ denotes the node at the bubble surface $\xi=\alpha$,
\begin{equation}
\widehat{{\bm e}_r\cdot{\bm e}_\xi}=
\frac{\delta_j(g_{\xi\eta}^{(\xi)})}{\delta_j(g_{\eta\phi})},
\ \ \ 
{\bm e}_z\cdot{\bm e}_\xi=
-\frac{\delta_j(r^2)}{2\delta_j(g_{\eta\phi})},
\end{equation}
\begin{equation}
g_{\xi\eta}^{(\xi)}
=\int_{\pi}^\eta\!\!\!{\rm d}\bar{\eta}
\left(-\frac{k^2{\cal S}\sin\bar{\eta}}{{\cal D}^3}\right)
=
k^2\Biggl\{
\frac{1}{\sinh^2\xi}
\tan^{-1}\left(\frac{{\cal D}+{\cal C}}{{\cal S}}\right)
+\frac{{\cal C}\sin\eta}{2{\cal D}^2\sinh\xi}
\Biggr\},
\end{equation}
\begin{equation}
\hat{\Sigma}_{\xi\eta}=
\frac{1}{\delta_i\delta_j(g_{\xi\eta})}
\left\{
\overline{\delta_i(g_\xi )}^j\delta_j(\hat{U}_\xi )-\delta_i\delta_j(g_\xi )\overline{\hat{U}_\xi }^j+
\overline{\delta_j(g_\eta)}^i\delta_i(\hat{U}_\eta)-\delta_i\delta_j(g_\eta)\overline{\hat{U}_\eta}^i
\right\},
\end{equation}
\begin{equation}
\hat{\Sigma}_{\xi\phi}=
\frac{1}{\delta_i(g_{\xi\phi})}
\left\{
\overline{r}^i\delta_i(\hat{U}_\phi)
-\delta_i(r)\overline{\hat{U}_\phi}^i
-\delta_i(g_\xi)\hat{U}_\xi
\right\},
\end{equation}
\begin{equation}
\hat{\sigma}_{\xi\eta}=
\frac{1}{\delta_i\delta_j(g_{\xi\eta})}
\left\{
\overline{\delta_i(g_\xi )}^j\delta_j(\hat{u}_\xi )-\delta_i\delta_j(g_\xi )\overline{\hat{u}_\xi }^j+
\overline{\delta_j(g_\eta)}^i\delta_i(\hat{u}_\eta)-\delta_i\delta_j(g_\eta)\overline{\hat{u}_\eta}^i
\right\},
\end{equation}
and the overline stands for the interpolation such as
\begin{equation}
\overline{q}^i|_{i,j}=\frac{q_{i+\frac{1}{2},j}+q_{i-\frac{1}{2},j}}{2},\ \ \ 
\overline{q}^j|_{i,j}=\frac{q_{i,j+\frac{1}{2}}+q_{i,j-\frac{1}{2}}}{2}.
\end{equation}
We write the area integral on the bubble surface in a summation form
\begin{equation}
\oint_{S_B}\!\!\!\!\!{\rm d}{\bm x}^2\ q
\equiv
\int_0^{2\pi}\!\!\!{\rm d}\phi
\int_0^{\pi}\!\!\!{\rm d}\eta\ h_\eta h_\phi\ q|_{\xi=\alpha}
=
\sum_{j=1}^{N_j} \left(\delta_j(g_{\eta\phi})
\int_0^{2\pi}\!\!\!{\rm d}\phi\ 
q\right)_{N_\alpha+\frac{1}{2},j},
\end{equation}
where $N_j$ is the number of grid points in the $\eta$-direction.
The drag force $F_{DC}$ in (\ref{eq:fdc}) is given by 
\begin{equation}
F_{DC}=
2\pi \sum_{j=1}^{N_j} \left(\delta_j(g_{\eta\phi})
{\bm e}_z\cdot{\bm e}_\xi
\overline{\hat{\sigma}_{\xi\xi}}^i
\right)_{N_\alpha+\frac{1}{2},j},
\end{equation}
where 
\begin{equation}
\hat{\sigma}_{\xi\xi}=
-\hat{p}+
\frac{2\overline{\delta_j(g_{\eta\phi})}^i\delta_i(\hat{u}_\xi)}{\delta_i\delta_j(g)}
+\frac{2\overline{r}^j\delta_i\delta_j(g_\xi)\overline{\hat{u}_\eta}^j}{\delta_i\delta_j(g)}.
\end{equation}
For the deflection $\hat{f}=f^{({\rm Ca})}/(a\cos\phi)$, 
the Laplace law (\ref{eq:laplace_law}) is expressed as
\begin{equation}
a_{n,j}^{(f)}\hat{f}|_{j+1}+a_{s,j}^{(f)}\hat{f}|_{j-1}-a_{p,j}^{(f)}\hat{f}|_{j}
=S^{(f)}|_j,
\end{equation}
where
\begin{equation}
\left.
\begin{array}{l}
\displaystyle
a_{n,j}^{(f)}=\frac{r_{N_\alpha+\frac{1}{2},j+\frac{1}{2}}}
{\overline{\delta_j(g_{\eta})}^i|_{N_\alpha+\frac{1}{2},j+\frac{1}{2}} 
\delta_j(g_{\eta\phi})|_{N_\alpha+\frac{1}{2},j}},\\
\displaystyle
a_{s,j}^{(f)}=\frac{r_{N_\alpha+\frac{1}{2},j-\frac{1}{2}}}
{\overline{\delta_j(g_{\eta})}^i|_{N_\alpha+\frac{1}{2},j-\frac{1}{2}} 
\delta_j(g_{\eta\phi})|_{N_\alpha+\frac{1}{2},j}},\\
\displaystyle
a_{p,j}^{(f)}=\frac{
a_{n,j}^{(f)}(\widehat{{\bm e}_r\!\cdot\!{\bm e}_\xi})_{N_\alpha+\frac{1}{2},j+1}+
a_{s,j}^{(f)}(\widehat{{\bm e}_r\!\cdot\!{\bm e}_\xi})_{N_\alpha+\frac{1}{2},j-1}
}{           (\widehat{{\bm e}_r\!\cdot\!{\bm e}_\xi})_{N_\alpha+\frac{1}{2},j  }},
\end{array}
\right\},
\end{equation}
\begin{equation}
S^{(f)}=-\overline{\hat{\Sigma}_{\xi\xi}}^i+\frac{
\widehat{{\bm e}_r\!\cdot\!{\bm e}_\xi}
\sum_{j=1}^{N_j}(\delta_j(g_{\eta\phi})\widehat{{\bm e}_r\!\cdot\!{\bm e}_\xi}\overline{\hat{\Sigma}_{\xi\xi}}^i)_j}{
\sum_{j=1}^{N_j}(\delta_j(g_{\eta\phi})\widehat{{\bm e}_r\!\cdot\!{\bm e}_\xi})^2_j},
\end{equation}
\begin{equation}
\hat{\Sigma}_{\xi\xi}=
-\hat{P}+
\frac{2\overline{\delta_j(g_{\eta\phi})}^i\delta_i(\hat{U}_\xi)}{\delta_i\delta_j(g)}
+\frac{2\overline{r}^j\delta_i\delta_j(g_\xi)\overline{\hat{U}_\eta}^j}{\delta_i\delta_j(g)}.
\end{equation}
The deformation-induced lateral force
$F_M^{({\rm Ca})}$ in (\ref{eq:fmca_st}) is given by 
\begin{equation}
\begin{split}
&F_M^{({\rm Ca})}
=
\pi\sum_{j}\biggl[
\frac{
\delta_{j}(g_{\eta\phi})
\overline{\hat{\sigma}_{\xi\xi}}^i
\delta_i(\hat{U}_\xi-\widehat{{\bm e}_r\!\cdot\!{\bm e}_\xi})
\hat{f}
}{\delta_i(g_\xi)}
+
\frac{
r\delta_i\delta_j(g_\xi)
\overline{\overline{\hat{\sigma}_{\xi\xi}}^i}^j
\overline{(\hat{U}_\eta-\widehat{{\bm e}_r\!\cdot\!{\bm e}_\eta})}^i
\overline{\hat{f}}^j
}{\overline{\delta_i(g_\xi)}^j}
\\&
+
\delta_j(g_{\eta\phi})\overline{\hat{\sigma}_{\xi\xi}}^i
\overline{\left(\frac{(\hat{U}_\phi+1)}{r}\right)}^i
\hat{f}
-
r\overline{\overline{\hat{\sigma}_{\xi\xi}}^i}^j
\overline{(\hat{U}_\eta-\widehat{{\bm e}_r\!\cdot\!{\bm e}_\eta})}^i
\delta_j(\hat{f})
\\&
-
\frac{
\overline{\delta_j(g_{\eta\phi})}^j
\overline{(\hat{u}_\eta+{\bm e}_z\!\cdot\!{\bm e}_\eta)}^i
\delta_i(\hat{\Sigma}_{\xi\eta})
\overline{\hat{f}}^j}{\delta_i(g_\xi)}
+
\frac{r
\delta_i\delta_j(g_\xi)
\overline{(\hat{u}_\eta+{\bm e}_z\!\cdot\!{\bm e}_\eta)}^i
}{\overline{\delta_i(g_\xi)}^j}
\left(
\overline{\overline{\hat{\Sigma}_{\xi \xi }}^i}^j-
\overline{\overline{\hat{\Sigma}_{\eta\eta}}^i}^j
\right)\overline{\hat{f}}^j
\\&
+
r
\overline{(\hat{u}_\eta+{\bm e}_z\!\cdot\!{\bm e}_\eta)}^i
\overline{\overline{\hat{\Sigma}_{\eta\eta}}^i}^j
\delta_j(\hat{f})
-
\frac{
\overline{(\hat{u}_\eta+{\bm e}_z\!\cdot\!{\bm e}_\eta)}^i
}{r}
\overline{
\left(
\delta_j(g_{\eta\phi})
\hat{\Sigma}_{\eta\phi}
\right)}^i
\overline{\hat{f}}^j
\\&
+
r\overline{(\hat{u}_\eta+{\bm e}_z\!\cdot\!{\bm e}_\eta)}^i
\overline{S^{(f)}}^j
\delta_j(\hat{f})
\biggr]_{N_\alpha+\frac{1}{2}},
\end{split}
\end{equation}
where
\begin{equation}
\widehat{{\bm e}_r\cdot{\bm e}_\eta}=\frac{\delta_i(g_{\xi\eta}^{(\eta)})}{\delta_i(g_{\xi\phi})},
\ \ \ 
{\bm e}_z\cdot{\bm e}_\eta=\frac{\delta_i(r^2)}{2\delta_i(g_{\xi\phi})},
\end{equation}
\begin{equation}
\begin{split}
g_{\xi\eta}^{(\eta)}
&=\int_{\alpha}^\xi\!\!\!{\rm d}\bar{\xi}\ 
\frac{k^2{\cal C}\sin\eta}{{\cal D}^3}
=
k^2\Biggl\{
-\frac{1}{\sin^2\eta}
\left(
\tan^{-1}\left(\frac{{\cal D}+{\cal C}}{{\cal S}}\right)-
\tan^{-1}\left(\frac{{\cal D}_\alpha+{\cal C}_\alpha}{{\cal S}_\alpha}\right)
\right)
\\&
-\left(
\frac{{\cal C}+2}{2{\cal S}}-\frac{{\cal C}_\alpha+2}{2{\cal S}_\alpha}
\right)
-\left(
\frac{{\cal C}\sin\eta}{2{\cal D}^2\sinh\xi}
-\frac{{\cal C}_\alpha\sin\eta}{2{\cal D}_\alpha^2\sinh\alpha}
\right)
\Biggr\},
\end{split}
\end{equation}
\begin{equation}
\hat{\Sigma}_{\eta\eta}=
-\hat{P}+
\frac{2\overline{\delta_i(g_{\xi\phi})}^j\delta_j(\hat{U}_\eta)}{\delta_i\delta_j(g)}
+\frac{2\overline{r}^i\delta_i\delta_j(g_\eta)\overline{\hat{U}_\xi}^i}{\delta_i\delta_j(g)},
\end{equation}
\begin{equation}
\hat{\Sigma}_{\eta\phi}=
\frac{1}{\delta_j(g_{\eta\phi})}
\left\{
\overline{r}^j\delta_j(\hat{U}_\phi)
-\delta_j(r)\overline{\hat{U}_\phi}^j
-\delta_j(g_\eta)\hat{U}_\eta
\right\}.
\end{equation}

\section{Small deformation theory in the lubrication limit}
\label{appendix_b}

The governing equations for
 $\hat{U}_r^{(0)}$, $\hat{U}_\phi^{(0)}$, $\hat{U}_z^{(0)}$, $\hat{P}^{(0)}$, 
 $\hat{U}_r^{(f)}$, $\hat{U}_z^{(f)}$ and $\hat{P}^{(f)}$ 
 in (\ref{eq:ur_lub})--(\ref{eq:def_f_lub}) are written as
\begin{equation}
\frac{1}{R}\frac{\partial(R \hat{U}_r^{(0)})}{\partial R}+
\frac{\hat{U}_\phi^{(0)}}{R}+
\frac{\partial \hat{U}_z^{(0)}}{\partial Z}=
\frac{1}{R}\frac{\partial(R \hat{U}_r^{(f)})}{\partial R}+
\frac{\partial \hat{U}_z^{(f)}}{\partial Z}=0,
\label{eq:cont}
\end{equation}
\begin{equation}
\begin{split}
&-\frac{\partial \hat{P}^{(0)}}{\partial R}+\frac{\partial^2 \hat{U}_r^{(0)}}{\partial Z^2}
=\frac{\hat{P}^{(0)}}{R}+\frac{\partial^2 \hat{U}_\phi^{(0)}}{\partial Z^2}
=\frac{\partial \hat{P}^{(0)}}{\partial Z}
\\=&
-\frac{\partial \hat{P}^{(f)}}{\partial R}+\frac{\partial^2 \hat{U}_r^{(f)}}{\partial Z^2}
=\frac{\partial \hat{P}^{(f)}}{\partial Z}
=0,
\end{split}
\label{eq:mom}
\end{equation}
with the no-slip boundary condition on the plane wall 
\begin{equation}
{\bm u}=0\ \ \ {\rm at}\ \ \ Z=0,
\label{eq:noslip_wall}
\end{equation}
and the free-slip and kinematic boundary conditions on the bubble surface.
The deformed interface is located on the curve 
where $H-\delta\hat{F}\cos\phi-Z=0$ holds.
From this relation, 
the normal unit vector ${\bm n}$ pointing outwards the liquid on the
bubble surface is approximated by
\begin{equation}
{\bm n}(R,\phi)=
-{\bm e}_z
+{\bm e}_r\varepsilon^{1/2}\left(
R-\delta\frac{{\rm d}\hat{F}}{{\rm d}R}\cos\phi
\right)
+{\bm e}_\phi \varepsilon^{1/2}\delta\frac{\hat{F}}{R}\sin\phi+...
\label{eq:approx_n}
\end{equation}
Applying the Taylor expansion to a function
 $q=q^{(0)}+\delta\ q^{(f)}+...$
 in terms of the deflection around the undeformed interface,
 one obtains a relation on the deformed interface $Z=H-\delta\hat{F}\cos\phi$
\begin{equation}
\begin{split}
q|_{\rm interface}
=&
q|_{Z=H}^{(0)}
+
\delta
\left(
-\hat{F}\cos\phi\left.
\frac{\partial q^{(0)}}{\partial Z}\right|_{Z=H}
+q|_{Z=H}^{(f)}
\right)+...
\label{eq:approx_q}
\end{split}
\end{equation}
Taking (\ref{eq:approx_n}) and (\ref{eq:approx_q}) into account, 
 one writes the kinematic condition on the bubble surface as
\begin{align}
&R\hat{U}_r^{(0)}-\hat{U}_z^{(0)}-R
\\=&
R\hat{U}_r^{(f)}-\hat{U}_z^{(f)}
-\frac{1}{2}\frac{{\rm d}\hat{F}}{{\rm d}R}\hat{U}_r^{(0)}
+\frac{\hat{F}}{2R}\hat{U}_\phi^{(0)}
-\frac{\hat{F}R}{2}\frac{\partial \hat{U}_r^{(0)}}{\partial Z}
\nonumber\\&
+\frac{\hat{F}}{2}\frac{\partial \hat{U}_z^{(0)}}{\partial Z}
+
\frac{1}{2}
\left(
\frac{{\rm d}\hat{F}}{{\rm d}R}
+\frac{\hat{F}}{R}
\right)=0
\ \ {\rm at}\ \ Z=H,
\label{eq:bc_kin1}
\end{align}
and the free-slip boundary condition as
\begin{align}
\frac{\partial \hat{U}_r^{(0)}}{\partial Z}=
\frac{\partial \hat{U}_\phi^{(0)}}{\partial Z}=
\frac{\partial \hat{U}_r^{(f)}}{\partial Z}
-\frac{\hat{F}}{2}\frac{\partial^2 \hat{U}_r^{(0)}}{\partial Z^2}=
0
\ \ \ {\rm at}\ \ \ Z=H.
\label{eq:bc_fsr0}
\end{align}
The vertical drag force acting on the bubble, 
 which is involved in the second term
 in the right-hand-side of (\ref{eq:laplace_law}),
 is of order 
$\log\varepsilon$ as determined
for a motion of a rigid sphere (\cite{gol1967, one1967})
 by means of the matched asymptotic expansion technique
 and the normal stress on the bubble surface 
 is dominated by the pressure $p = O(\varepsilon^{-3/2})$
 as compared with 
$\partial_r u_r = O(\varepsilon^{-1/2})$, 
$u_r/r = O(\varepsilon^{-1/2})$, 
$u_\phi/r = O(\varepsilon^{-1/2})$ and 
$\partial_z u_z = O(\varepsilon^{-1/2})$, 
which are related to the viscous stresses.
Hence, the Laplace law (\ref{eq:laplace_law}) is simplified into 
\begin{equation}
\frac{{\rm d}}{{\rm d}R}\left(
\frac{1}{R}
\frac{{\rm d}(R\hat{F})}{{\rm d}R}
\right)=\hat{P}^{(0)},
\label{eq:laplace1}
\end{equation}
with no singularity conditions $\hat{F}=0$ at $R=0$
and $\hat{F}\rightarrow 0$ as $R\rightarrow \infty$.

{As obtained by \cite{hod2004}, the leading-order and perturbed solutions are}
\begin{equation}
\begin{split}
&\hat{U}_r^{(0)}=\frac{(6-9R^2)}{20H}\left(
\frac{Z^2}{H^2}-\frac{2Z}{H}\right),
\ \ \ 
\hat{U}_\phi^{(0)}=-\frac{3}{10}\left(
\frac{Z^2}{H^2}-\frac{2Z}{H}\right),
\\&
\hat{U}_z^{(0)}=
\frac{(4R-R^3)Z^3}{5H^4}+\frac{(-42R+3R^3)Z^2}{20H^3},
\label{eq:u_lub}
\end{split}
\end{equation}
\begin{equation}
\hat{P}^{(0)}=\frac{3R}{5H^2},
\label{eq:p_lub}
\end{equation}
\begin{equation}
\hat{F}=-\frac{3\log H}{5R}.
\label{eq:f_lub}
\end{equation}
\begin{equation}
\begin{split}
\hat{U}_r^{(f)}(R,Z)=&
A_2(R)Z^2+A_1(R)Z,\\
\hat{U}_z^{(f)}(R,Z)=&
-\frac{1}{3R}\frac{{\rm d}(RA_2)}{{\rm d}R}Z^3
-\frac{1}{2R}\frac{{\rm d}(RA_1)}{{\rm d}R}Z^2,
\label{eq:uz1}
\end{split}
\end{equation}
\begin{equation}
\hat{P}^{(f)}(R)=\int_\infty^R\!\!\!{\rm d}\bar{R}\ 
2A_2(\bar{R}),
\label{eq:p1}
\end{equation}
where
\begin{equation}
A_1=\frac{9(14-R^2)\log H}{100H^3R},
\label{eq:a1}
\end{equation}
\begin{equation}
A_2=-\frac{9(4-R^2)\log H}{50H^4R}.
\label{eq:a2}
\end{equation}

Substituting (\ref{eq:a2}) into (\ref{eq:p1}) estimates the asymptotic order of the perturbed pressure
$$
\varepsilon^{-3/2}\delta
\hat{P}^{(f)}=
\frac{12\varepsilon^{-3}{\rm Ca}
\log(R^2/2)}{25R^6}+O(R^{-6})
\ \ \ {\rm for}\ \ \ R\gg 1,
$$
which is $O({\rm Ca}\log\varepsilon\ (\varepsilon^{1/2}R)^{-6})$ 
in the overlapping region $R\sim \varepsilon^{-1/2}$, 
and thus to be matched with the pressure of 
$O({\rm Ca}\log\varepsilon\ r^{-6})$ in the outer region
(\cite{one1967}).
This outer pressure may contribute to 
the lateral force of $O({\rm Ca}\log\varepsilon)$,
which is larger than that for the wide-gap case, 
corresponding to $O({\rm Ca})$.
Nevertheless, as discussed in \cite{urz2007}, 
the rapid decay of the perturbed pressure for $R\gg 1$ 
indicates that the contribution of the outer pressure to the lateral
force is negligibly smaller than that of the inner pressure.
Therefore, 
 to evaluate the leading-order migration force, 
 one does not have to solve the outer problem. 
(In fact, the order of the lateral force (\ref{eq:fm_lub}) evaluated only in the
inner region is confirmed to be $O({\rm Ca}\varepsilon^{-2})$
and larger than the outer contribution $O({\rm Ca}\log\varepsilon)$.) 
The leading pressure $\varepsilon^{-3/2}\hat{P}^{(0)}\cos\phi$ with no deformation
is locally dominant but does not contribute to the lateral force 
due to its azimuthal cosine dependence. 
The viscous stress contribution to the lateral force is 
 $O(\varepsilon)$ smaller than the pressure in the inner region.
The deformation-induced lateral force $F_M$ to cancel the migration velocity 
and to maintain the wall-parallel motion is expressed as
\begin{equation}
F_M\approx 
\oint_{\rm contact}\!\!\!\!\!\!\!\!\!\!{\rm d}^2{\bm x}\ 
\varepsilon^{-3/2}\delta\hat{P}^{(f)}.
\label{eq:intp01}
\end{equation}
The surface integral for a function $q$ on the contact side
is taken from the axis $R=0$ to the overlapping region, i.e., 
\begin{equation}
\oint_{\rm contact}\!\!\!\!\!\!\!\!\!\!{\rm d}^2{\bm x}\ q
=\varepsilon \int_0^{2\pi}\!\!\!{\rm d}\phi\int_0^{\frak R}\!\!\!{\rm d}R\ R\ q(R,\phi),
\label{eq:surf_int_contact}
\end{equation}
where ${\frak R}=O(\varepsilon^{-1/2})$. 
For ${\frak R} \gg 1$, one obtains an asymptotic relation
\begin{equation}
\begin{split}
\int_0^{\frak R}\!\!\!\!\!\!{\rm d}R\ R\ \hat{P}^{(f)}(R)
=&
-\int_0^{\frak R}\!\!\!\!\!\!{\rm d}R\ R^2 A_2(R)
-{\frak R}^2\int_{\frak R}^{\infty}\!\!\!\!\!\!{\rm d}R\ A_2(R)
\\=&
\frac{3}{100}+\frac{6\log({\frak R}^2/2)}{25{\frak R}^4}
+\frac{1}{5{\frak R}^4}+O({\frak R}^{-6}\log {\frak R}),
\label{eq:int_pf_lar}
\end{split}
\end{equation}
of which only the first term does not vanish as ${\frak R}\rightarrow \infty$.
As obtained by \cite{hod2004},
 consequently,
 the asymptotic solution of the deformation-induced lateral 
 force in the lubrication limit is 
\begin{equation}
\begin{split}
F_M\rightarrow&
\lim_{{\frak R}\rightarrow\infty}
\varepsilon
\int_0^{2\pi}\!\!\!\!\!\!{\rm d}\phi
\int_0^{\frak R}\!\!\!\!\!\!{\rm d}R\ R\ 
\varepsilon^{-3/2}\delta\hat{P}^{(f)}(R)
\\=&
\frac{3\pi{\rm Ca}}{50}\varepsilon^{-2}
\ \ {\rm as}\ \ \varepsilon\rightarrow 0.
\label{eq:fm_lub}
\end{split}
\end{equation}

\end{appendix}

\end{document}